\documentclass[AMA,STIX1COL]{WileyNJD-v2}

\articletype{RESEARCH ARTICLE}

\usepackage{subfigure}
\usepackage{threeparttable}
\usepackage{graphicx}
\usepackage{hyperref}
\hypersetup{colorlinks = false}
\usepackage{float}

\graphicspath{{./picture/}}

\begin{document}

\title{{L}ife-cycle assessment for flutter probability of a long-span suspension bridge based on field monitoring data}

\author[1]{Xiaolei Chu}

\author[2]{Hung Nguyen Sinh}

\author[1,3]{Wei Cui*}

\author[1,3,4]{Lin Zhao}

\author[1,3]{Yaojun Ge}

\authormark{Chu \textsc{et al}}

\address[1]{\orgdiv{State Key Lab of Disaster Reduction in Civil Engineering}, \orgname{Tongji University}, \orgaddress{\state{Shanghai}, \country{China}}}

\address[2]{\orgname{Amazon.com Inc.}, \orgaddress{\state{Seattle, WA}, \country{USA}}}

\address[3]{\orgdiv{Key Laboratory of Transport Industry of Bridge Wind Resistance Technologies}, \orgname{Tongji University}, \orgaddress{\state{Shanghai}, \country{China}}}

\address[4]{\orgdiv{State Key Laboratory of Mountain Bridge and Tunnel Engineering}, \orgname{Chongqing Jiaotong University}, \orgaddress{\state{Chongqing}, \country{China}}}

\corres{*Wei Cui, 207 Wind Engineering Building, Tongji University, 1239 Siping Road, Shanghai, 200092, China. \email{cuiwei@tongji.edu.cn}}


\abstract[Summary]
{Assessment of structural safety status is of paramount importance for existing bridges, where accurate evaluation of flutter probability is essential for long-span bridges. In current engineering practice, at the design stage, flutter critical wind speed is usually estimated by the wind tunnel test, which is sensitive to modal frequencies and damping ratios. After construction, structural properties of existing structures will change with time due to various factors, such as structural deteriorations and periodic environments. The structural dynamic properties, such as modal frequencies and damping ratios, cannot be considered as the same values as the initial ones, and the deteriorations should be included when estimating the life-cycle flutter probability. This paper proposes an evaluation framework to assess the life-cycle flutter probability of long-span bridges considering the deteriorations of structural properties, based on field monitoring data. The Bayesian approach is employed for modal identification of a suspension bridge with the main span of 1650 m, and the field monitoring data during 2010-2015 is analyzed to determine the deterioration functions of modal frequencies and damping ratios, as well as their inter-seasonal fluctuations. According to the historical trend, the long-term structural properties can be predicted, and the probability distributions of flutter critical wind speed for each year in the long term are calculated. Consequently, the life-cycle flutter probability is estimated, based on the predicted modal frequencies and damping ratios.}



\maketitle


\section{Introduction}\label{sec:introduction}

Flutter is a dynamic instability phenomenon of an elastic structure in the wind flow.
In 1940, Tacoma Narrows Bridge collapsed four months after it was built, attracting the attention of flutter reliability. 
Severe consequences induced by flutter have motivated investigations for prediction of the flutter critical wind speed \cite{theodorsen1949general,jones2001theory} and for assessment of flutter probability \cite{seo2011estimation,ji2020probabilistic,fang2020experimental}.
Furthermore, with the ever-growing increase of bridge span, assessment for aerodynamic performance becomes much more significant.

Modal frequencies and damping ratios are two important parameters for bridge flutter-resistance performance. In existing methods, the evaluation of flutter performance for long-span bridges has been comprehensively conducted by wind tunnel tests \cite{yang2015aerodynamic,yang2015aerodynamic2} and numerical methods \cite{caracoglia2011simulation,yang2007investigation}. However, the investigation of life-cycle flutter probability of in-service long-span bridges is very rare, especially considering structural deteriorations of dynamic properties based on field monitoring data. In recent years, structural health monitoring (SHM) system, as an effective tool to record structural responses \cite{ni2009technology}, can provide sufficient structural vibration information and external excitation, and infer structural conditions \cite{li2014smc}. Based on these data, it is possible to analyze the structural dynamic properties and corresponding changing trends comprehensively, and predict their future performance consequently.

Practically, ambient vibration tests have attracted increasing attention in modal identification, which can be performed effectively onto structures in working conditions and without artificial intervention \cite{katafygiotis2001bayesian,au2012fastpart1,au2012fastpart2}. Ambient modal identification, including the determination of modal frequencies, damping ratios, and mode shapes of a constructed structure using measured data \cite{ewins1984modal}, does not require loading conditions in advance, but assume the external excitation is statistically random rather than constant. Stochastic subspace identification (SSI) method \cite{peeters2001stochastic} and Enhanced Frequency Domain Decomposition (EFDD) method \cite{brincker2001modal} in time domain are popular techniques capable of quickly extracting structural properties. Bayesian system identification approach \cite{beck1998updating,beck2010bayesian,zhang2017bayesian} is another popular method, viewing modal identification as an inference problem where probability is utilized as a measure for the plausibility of outcomes given a model of the system and measured data. The general principle of Bayesian identification is considering the objectives as a joint posterior probability density function (PDF) of the modal parameters for given measured data and modeling assumptions. In typical applications where there is a sufficiently large amount of data, the modal parameters are ``globally identifiable'' and the posterior PDF can be well approximated by a Gaussian PDF \cite{yuen2010bayesian}, which is completely characterized by the most probable value (MPV) and the covariance matrix \cite{au2012fastpart1,au2012fastpart2,yuen2010bayesian}. 

On the other hand, the assessment of flutter-resistance performance is an important research field for bridge engineering, considering the emerging trend of performance-based wind engineering \cite{ciampoli2011performance,Cui201875,spence2014performance}. Current reliability evaluation methods for bridge flutter-resistance performance  \cite{seo2015exploring,ge2000application,prenninger1990reliability} are based on multi-mode  \cite{jones2001theory} and full-mode \cite{ge2000aerodynamic} approaches, and incorporate the effects of uncertainty either by numerical simulations \cite{seo2015exploring} (e.g., Monte-Carlo sampling method) or by probability propagation methods \cite{mannini2015aerodynamic,Cui2015183}. Recent studies in existing literatures have suggested that the flutter probability can be successfully assessed by the first-order reliability method (FORM) \cite{dragomirescu2003probabilistic}, the ``response surface method'' \cite{cheng2005flutter}, the polynomial chaos expansion \cite{rizzo2018examination} and the artificial neural network model \cite{rizzo2020artificial}, etc.. Another method \cite{pourzeynali2002reliability} has proposed the solution to the flutter probability problem by perturbation of the deterministic flutter velocity through a set of dimensionless multiplicative random parameters. Canor and Caracoglia \cite{canor2015application} discussed the advantages and limitations of several reliability-based methods (e.g., random perturbation analysis, collocation methods, Galerkin approach). In this paper, a linear regression model is proposed to derive the PDF of flutter critical wind speed directly from the PDFs of modal frequencies and damping ratios. Long-span bridges are vulnerable to long-term environmental corrosion and fatigue damage accumulation, which is stochastic in nature and makes structural reliability time-variant. The challenges posed by aging structures have prompted several research programs to address risk management problems \cite{petcherdchoo2008optimizing,vu2000structural,mori1993reliability}. Assessment procedures considering modification of structural properties can have a better insight into structural performances over extended time frames \cite{ellingwood2016life,lee2017decision}.

In this paper, an evaluation framework is proposed to assess life-cycle reliability of existing bridges in terms of flutter-resistance performance, based on field monitoring data. The historical trends are fitted to modify structural properties, utilizing monitoring data from 2010 to 2015. Furthermore, modal frequencies and damping ratios are sensitive to the in-service environment \cite{alampalli1998influence}, and the annual fluctuations of modal properties due to uncertain environmental factors are considered to obey time-invariant probability distributions fitted by field monitoring results. At last, probability distributions of flutter critical wind speed are calculated for each year, then the time-variant flutter probability is subsequently predicted.

\section{Background methodology} 
\label{sec:background_theory_7}
\subsection{Flutter critical wind speed} 
\label{sec:flutter critical wind speed}
\begin{figure}[!htb]
    \centering
    \includegraphics[]{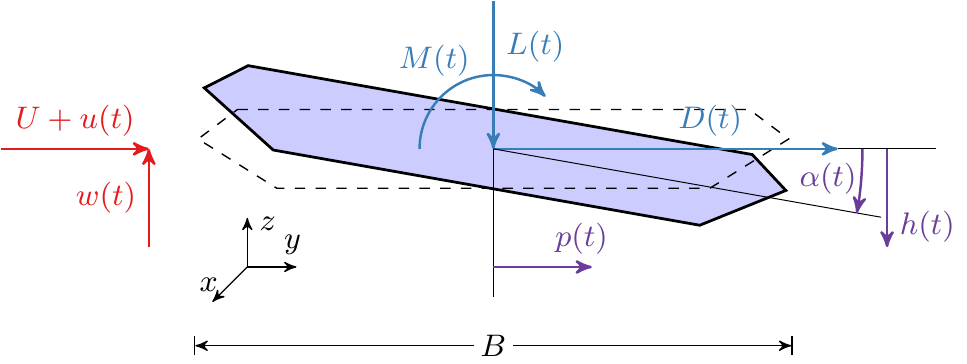}
    \caption{Bridge section and coordinate definition}
	\label{fig:bridge_section}
\end{figure}

In multi-mode analysis, deflection components of the bridge deck are represented in terms of generalized coordinate $\xi_{i}(t_{s})$, deck width $B$, and dimensionless modal values of the $i$-th mode along the deck $h_{i}(x)$, $p_{i}(x)$, and $\alpha_{i}(x)$, as shown in Fig.~\ref{fig:bridge_section}.

Bridge aeroelastic forces can be simplified by exclusively considering the vertical force (${L}_{ae}$) and torsional moment (${M}_{ae}$) per unit deck span \cite{scanlan1971air}, which are expressed by

\begin{subequations}
\label{eq:aeroelastic force}
\begin{align}
L_{\mathrm{ae}}(x, t_{s}) =\frac{\rho U^{2} B}{2}\left[\begin{array}{l}
K H_{1}^{*} \dot{h}(x, t_{s}) / U+K H_{2}^{*} B \dot{\alpha}(x, t_{s}) / U +\\
K^{2} H_{3}^{*} \alpha(x, t_{s})+K^{2} H_{4}^{*} h(x, t_{s}) / B
\end{array}\right] \\
M_{\mathrm{ae}}(x, t_{s}) =\frac{\rho U^{2} B^{2}}{2}\left[\begin{array}{l}
K A_{1}^{*} \dot{h}(x, t_{s}) / U+K A_{2}^{*} B \dot{\alpha}(x, t_{s}) / U +\\
K^{2} A_{3}^{*} \alpha(x, t_{s})+K^{2} A_{4}^{*} h(x, t_{s}) / B
\end{array}\right],
\end{align}
\end{subequations}
where $B$ is the deck width; $U$ is the mean wind velocity at the deck level, acting orthogonally to the longitudinal axis; $\rho$ is the air density; $t_{s}$ is the time variable, unit of which is second; $K = \omega B / U$ is the reduced frequency; $\dot{h}(t_{s}) = \mathrm{d} h / \mathrm{d} t_{s}$ and $\dot{\alpha}(t_{s}) = \mathrm{d} \alpha / \mathrm{d} t_{s}$. By modal superpositions, the dynamic response in Fig.~\ref{fig:bridge_section} of can be expressed as

\begin{subequations}
\label{eq:h_alpha}
\begin{align}
h(x, t_{s}) &=\sum_{g} h_{g}(x) B \xi_{g}(t_{s})\\
\alpha(x, t_{s}) &=\sum_{g} \alpha_{g}(x) \xi_{g}(t_{s})
\end{align}
\end{subequations}
where $x$ is the coordinate along the deck span; $t_{s}$ is the time; $h_{g}(x)$ and $\alpha_{g}(x)$ are the dimensionless $g$-th mode shapes; $\xi_{g}(t_{s})$ are associated generalized coordinates.

A coupled-two-mode system of dynamic equations can be further simplified by the first vertical mode and first torsional one as "$v1$" and "$t1$" with circular frequencies $\omega_{v}$ and $\omega_{t1}$. The corresponding expressions are also simplified as uncoupled mode shapes $h_{v1}(x)$ and $\alpha_{v2}(x)$, respectively. Then Eq.~\eqref{eq:h_alpha} can be simplified as

\begin{subequations}
\label{eq:h_alpha_simplified}
\begin{align}
h(x, t_{s}) & \cong \xi_{v1}(t_{s}) B h_{v1}(x) \\
\alpha(x, t_{s}) & \cong \xi_{t1}(t_{s}) \alpha_{t1}(x)
\end{align}
\end{subequations}

The coupled-mode aeroelastic instability can be derived \cite{caracoglia2009comparative} by Fourier-domain analysis of two-dimensional system of equations described above in terms of complex amplitudes of the generalized $\xi_{v1}$, $\xi_{t1}$ in Eq.~\eqref{eq:h_alpha_simplified}, expressed in vector form as $\bar{\xi}=\left[\bar{\xi}_{v 1}, \bar{\xi}_{t 1}\right]^{T}$. On-going flutter can be determined after representing the simple harmonic motion for both modes in terms of a critical reduced frequency ratio $\chi=K / K_{t 1}$ \cite{caracoglia2009comparative}. The problem can be reduced to the nontrivial solutions of a two-by-two complex algebraic system $\mathbf{E}(K, \chi) \bar{\xi}=\mathbf{0}$ with $q_{v1} = 0.5 \rho B^{4} \ell / I_{v1}$ and $q_{t1} = 0.5 \rho B^{4} \ell / I_{t1}$. The scalar components of $\mathbf{E}(K, \chi)$ with $\mathbf{i} = \sqrt{-1}$ are 

\begin{subequations}
\label{eq:E_ij}
\begin{align}
E_{1,1}(K, \chi) & =\left[\begin{array}{c}
-\chi^{2}+\left(K_{v 1} / K_{t 1}\right)^{2}-q_{v 1} \chi^{2} H_{4}^{*}(K) G_{v 1, v 1} \\
+\mathbf{i}\left(2 \zeta_{v 1} K_{v 1} / K_{t 1} \chi-q_{v 1} \chi^{2} H_{1}^{*}(K) G_{v 1, v 1}\right)
\end{array}\right] \\
E_{1,2}(K, \chi) &=-q_{v 1}\left(\chi^{2} H_{3}^{*}(K) G_{v 1, t 1}+\mathbf{i} \chi^{2} H_{2}^{*}(K) G_{v 1, t 1}\right) \\
E_{2,1}(K, \chi) &=-q_{v 1}\left(\chi^{2} A_{4}^{*}(K) G_{v 1, t 1}+\mathbf{i} \chi^{2} A_{1}^{*}(K) G_{v 1, t 1}\right) \\
E_{2,2}(K, \chi) & =\left[\begin{array}{c}
\chi^{2}+1-q_{t 1} \chi^{2} A_{3}^{*}(K) G_{t 1, t} \\
+\mathbf{i}\left(2 \zeta_{t 1} \chi-q_{t 1} \chi^{2} A_{2}^{*}(K) G_{t 1, t 1}\right)
\end{array}\right]
\end{align}
\end{subequations}
where $\zeta_{v1}$ and $\zeta_{t1}$ are modal damping ratios; the reduced frequencies of the modes are $K_{v1} = \omega_{v1} B /U$ and $K_{t1} = \omega_{t1} B / U$. The dimensionless modal integrals \cite{caracoglia2009comparative} in Eq.~\eqref{eq:E_ij} are : $G_{v 1, v 1}=\int_{0}^{\ell} h_{v 1}^{2}(x) \mathrm{d} x / \ell$, $G_{t 1, t 1}=\int_{0}^{\ell} \alpha_{t 1}^{2}(x) \mathrm{d} x / \ell$ and $G_{v 1, t 1}=\int_{0}^{\ell} h_{v 1}(x) \alpha_{t1}(x) \mathrm{d} x / \ell$ with $\ell$ being the longitudinal length of the bridge. Generalized modal inertias are $I_{v1} = m_{0} \ell B^{2} G_{v1, v1}$ and $I_{t1} = I_{0} \ell G_{t1,t1}$, with $m_{0}$ and $I_{0}$ being an equivalent mass and a mass moment of inertia perunit length of the moving structure. An iterative procedure is needed to solve for $\operatorname{det}[\mathbf{E}(K, \chi)]=0$ \cite{caracoglia2009comparative}.

\subsection{Bayesian FFT Modal Identification} 
Yuen, Katafygiotis \cite{yuen2003bayesian} and Au et al.\cite{au2012fastpart1,au2012fastpart2} developed an efficient algorithm to identify most probable values (MPV) and corresponding covariances of modal properties (e.g., modal frequencies and damping ratios) with well-separated modes and possibly close-spaced modes \cite{brownjohn2018bayesian}. The definitions of some quantities in this section have been adapted from the original formulation of Au \cite{au2012fastpart1}. If the acceleration time history measured at $n$ DOFs of a structure is noted as \{$\hat{\mathbf{x}}_{j} \in R^{n}: j=1, \ldots, N$\} and abbreviated as \{$\hat{\mathbf{x}}_{j}$\}, where $N$ = number of samples per channel, the FFT of \{$\hat{\mathbf{x}}_{j}$\} is defined as
\begin{equation}\label{eq:F_k}
\mathcal{F}_{k} =\mathbf{F}_{k}+\mathbf{i} \mathbf{G}_{k} =\sqrt{(2 \Delta t) / N} \sum_{j=1}^{N} \hat{\mathbf{x}}_{j} \exp \{-2 \pi \mathbf{i}[(k-1)(j-1) / \mathrm{N}]\} \ (k=1, \ldots, N)
\end{equation}
where $\mathbf{i}^{2} = -1$; $\Delta$t is the sampling interval; $\mathbf{F}_{k}=\operatorname{Re} \mathcal{F}_{k}$ and $\mathbf{G}_{k}=\operatorname{Im} \mathcal{F}_{k}$ denote the real and imaginary part of the FFT, respectively. For $k = 2,3, \dots,N_{q}$, the FFT corresponds to frequency $f_{k} = (k-1) / N \Delta{t}$. Here, $N_{q}$ = int[N / 2] + 1 (int
[$\cdot$] denotes the integer part) corresponds to the FFT ordinate at the Nyquist frequency. For modal identification only these ($N_{q} - 1$) FFT values are utilized.

In the context of Bayesian inference, the measured acceleration is modeled as $\hat{\mathbf{x}}_{j}=\mathbf{x}_{j}(\boldsymbol{\theta})+\mathbf{\epsilon}_{j}$ where $\mathbf{x}_{j}(\boldsymbol{\theta})$ is the acceleration  response of the structural model defined by the set of model parameters $\boldsymbol{\theta}$, the subject to be identified; $\epsilon_{j}$ is the prediction error that accounts for the deviation between the model response and measured data, possibly owing to measurement noise and modeling error. 
Yuen and Katafygiotis derived the joint PDF for the augmented FFT vectors \{$\mathbf{Z}_{k} = \left[\mathbf{F}_{k}^{T}, \mathbf{G}_{k}^{T}\right]^{T} \in R^{2 n}: k=2, \dots, N_{q}$\} and applied it to Bayesian modal identification \cite{yuen2003bayesian}. For a high sampling rate and long duration of data, $\mathbf{Z}_{k}$ is a zero-mean Gaussian vector with covariance matrix given by
\begin{equation}\label{eq:C_k}
\mathbf{C}_{k}=\frac{1}{2}\left[\begin{array}{ll}
	{\mathbf{\Phi}\left(\operatorname{Re} \mathbf{H}_{k}\right) \mathbf{\Phi}^{T}} & -{\mathbf{\Phi}\left(\operatorname{Im} \mathbf{H}_{k}\right) \mathbf{\Phi}^{T}} \\
{\mathbf{\Phi}\left(\operatorname{Im} \mathbf{H}_{k}\right) \mathbf{\Phi}^{T}} & {\mathbf{\Phi}\left(\operatorname{Re} \mathbf{H}_{k}\right) \mathbf{\Phi}^{T}}
\end{array}\right]+\left(\sigma^{2} / 2\right) \mathbf{I}_{2 n}
\end{equation}
where $\mathbf{\Phi} \in R^{n \times m}$ is the mode shape matrix confined to the measured DOFs (the $i$th column gives the $i$th mode shape); $\sigma ^{2}$ is the (constant) spectral density level of the prediction error; $\mathbf{I}_{2 n}$ denotes the $2 n \times 2 n$ identity matrix; $\mathbf{H}_{k}$ is the spectral density matrix of the model response and its ($i,j$) entry is given by 
\begin{equation}
\mathbf{H}_{k}(i, j)=S_{i j}\left[\left(\beta_{i k}^{2}-1\right)+\mathbf{i}\left(2 \zeta_{i} \beta_{i k}\right)\right]^{-1}\left[\left(\beta_{j k}^{2}-1\right)-\mathbf{i}\left(2 \zeta_{j} \beta_{j k}\right)\right]^{-1}
\end{equation}
where $\beta_{i k}=f^{(i)} / f_{k}$ = frequency ratio; $f^{(i)}$ and $\zeta_{i}$ = natural frequency and damping ratio of the $i$th mode, respectively; $S_{i j}$ = cross spectral density between the $i$th and $j$th modal excitation. 

The set of modal parameters ${\boldsymbol{\theta}}$ consists of modal frequencies, damping ratios, mode shapes, entries \{$\mathbf{S}_{i j}$\} of the spectral densty matrix of modal excitations and spectral density $\sigma ^{2}$ of the prediction error. Assuming a noninformative prior distribution, the posterior PDF of $\boldsymbol{\theta}$ given the FFT data is proportional to the likelihood function $p\left(\left\{\mathbf{Z}_{k}\right\} | \boldsymbol{\theta}\right)$
\begin{equation}
\begin{aligned}
p\left(\boldsymbol{\theta} |\left\{\mathbf{Z}_{k}\right\}\right) & \propto p\left(\left\{\mathbf{Z}_{k}\right\} | \boldsymbol{\theta}\right)=(2 \pi)^{-\left(N_{q}-1\right) / 2}\left[\prod_{k=2}^{N_{q}} \operatorname{det} \mathbf{C}_{k}(\boldsymbol{\theta})\right]^{-1 / 2} \\
& \times \exp \left[-(1 / 2) \sum_{k=2}^{N_{q}} \mathbf{Z}_{k}^{T} \mathbf{C}_{k}(\boldsymbol{\theta})^{-1} \mathbf{Z}_{k}\right]
\end{aligned}
\end{equation}
where the dependence of $\mathbf{C}_{k}$ on $\boldsymbol{\theta}$ has been emphasized \cite{yuen2010bayesian}. It is convenient to write with the negative log-likelihood function $L(\boldsymbol{\theta})$ form
\begin{equation}\label{eq:p_theta}
p\left(\boldsymbol{\theta} |\left\{\mathbf{Z}_{k}\right\}\right) \propto \exp [-L(\boldsymbol{\theta})]
\end{equation}
where
\begin{equation}\label{eq:L_theta}
\begin{aligned}
L(\boldsymbol{\theta}) &=(1 / 2) \sum_{k=2}^{N_{q}}\left[\ln \operatorname{det} \mathbf{C}_{k}(\boldsymbol{\theta})+\mathbf{Z}_{k}^{T} \mathbf{C}_{k}(\boldsymbol{\theta})^{-1} \mathbf{Z}_{k}\right] \\
&\approx L(\hat{\boldsymbol{\theta}})+\frac{1}{2}(\boldsymbol{\theta}-\hat{\boldsymbol{\theta}})^{T} H_{L}(\hat{\boldsymbol{\theta}})(\boldsymbol{\theta}-\hat{\boldsymbol{\theta}})
\end{aligned}
\end{equation}
where the first-order term vanishes owing to optimality of $\hat{\boldsymbol{\theta}}$; $H_{L}(\hat{\boldsymbol{\theta}})$ is the Hessian of $L$ at MPV (most probable value) \cite{au2012fastpart1,au2012fastpart2}. Substituting into Eq.~\eqref{eq:p_theta}, the posterior PDF becomes a Gaussian PDF
\begin{equation}
p\left(\boldsymbol{\theta} |\left\{\mathbf{Z}_{k}\right\}\right) \propto \exp \left[-(1 / 2)(\boldsymbol{\theta}-\hat{\boldsymbol{\theta}})^{T} \hat{\mathbf{C}}^{-1}(\boldsymbol{\theta}-\hat{\boldsymbol{\theta}})\right]
\end{equation}
where
\begin{equation}
\hat{\mathbf{C}}=H_{L}(\hat{\boldsymbol{\theta}})^{-1}
\end{equation}
is the posterior covariance matrix.

MPVs of modal properties can be determined by minimizing $L(\boldsymbol{\theta})$. Practically, computational problems are inevitable while identifying mode shapes due to the large number of measured DOFs and calculating the inverse of the covariance matrix $\mathbf{C}_{k}$, which have been well explained and solved in \cite{au2012fastpart1,au2012fastpart2,zhang2013erratum}.

\subsection{Life-cycle assessment of flutter reliability} 
Long-span bridges are susceptible to aging from various environmenal or non-environmental factors, such as chemical attack, corrosion, climate change and other physical mechanisms \cite{ellingwood2016life,lee2017decision}. Age-related deterioration leads to a decrease of structural capacity to withstand various challenges during its service life from operating conditions, natural environments, and accidents \cite{frangopol1997life,ellingwood2016life}, which makes life-cycle assessment necessary. Life-cycle assessment of long-span bridges ought to take possible deteriorations into accounts. SHM system offers a chance to monitor structural long-term structural properties, then corresponding changing trends can be found. 

As for flutter-resistance performance, factors that affect the flutter stability can be summarized into two aspects, structural flutter-resistance ability $V_{R}$ (i.e., flutter critical wind speed) and wind load effect $V_{S}$ (i.e., site wind speed). In flutter analysis, $V_{R}$ is the function of modal frequencies $\boldsymbol{f}$, damping ratios $\boldsymbol{\zeta}$ and flutter derivatives $\mathbf{A}^{*}$ ($A_{1}^{*}$, $A_{2}^{*}$, $A_{3}^{*}$, $A_{4}^{*}$), $\mathbf{H}^{*}$ ($H_{1}^{*}$, $H_{2}^{*}$, $H_{3}^{*}$, $H_{4}^{*}$). Structural properties will be time-variant due to aging and environmental factors. Based on field monitoring, changing trends of modal frequencies and damping ratios can be investigated. $V_{S}$ is the site wind speed, which is normally described as extreme wind speeds distribution for certain time duration. For example, in the wind load design standard \cite{chen2018wind}, Gumbel distribuiton is used to model annual extreme wind speeds. Based on $V_{R}$ and $V_{S}$, a state function $Z$ can be defined
\begin{equation}
Z\left(V_{R}, V_{S} ; t\right)=V_{R}\left(\boldsymbol{f}(t), \boldsymbol{\zeta}(t), \mathbf{A}^{*}, \mathbf{H}^{*}\right)-V_{S}(t)
\end{equation}
where $Z>0$ denotes structural safety, $Z=0$ denotes limit state and $Z<0$ denotes structural failure; $t$ is the time variable, unit of which is year.

In this research, $\boldsymbol{f}(t)$ and $\boldsymbol{\zeta}(t)$ are time-variant modal frequencies and damping ratios, respectively, fitted by field monitoring data; $\mathbf{A}^{*}$ and $\mathbf{H}^{*}$ are flutter derivatives, which are obtained by wind tunnel test and are functions of reduced velocities; $V_{S}(t)$ is the time-variant site wind speed. In reliability analysis, normal distribution is usually regarded as the most widely-used and accepted distribution for flutter derivatives considering its flexibility and simplicity \cite{mannini2015aerodynamic}. In this paper, the uncertainties of flutter derivatives are ignored because they are mainly caused by experimental error, and deteriorations due to aging are mainly concerned issues here. On the other hand, $V_{S}(t)$ is a time-variant variable susceptible to climate change, as preliminarily examined in \cite{seo2015exploring,cui2016exploring}. In this research, $V_{S}(t)$ is temporarily regarded to obey a constant Gumbel distribution, and its relationship with climate change remains to be investigated in the future. The PDFs of $V_{R}$ and $V_{S}$ are denoted as $f_{R}(r;t)$ and $f_{S}(s)$, respectively. Thus the structural failure (flutter on-site) probability \cite{melchers2018structural} is
\begin{equation}
\label{eq:PIM}
\begin{aligned}
P_{f}(t) &=P(Z \leqslant 0;t)=P(V_{R}(t)- V_{S} \leqslant 0) =\iint_{r \leqslant s} f_{R S}(r, s;t) \mathrm{d} r \mathrm{d} s \\
&=\iint_{r \leqslant s} f_{R}(r;t) \cdot f_{S}(s) \mathrm{d} r \mathrm{d} s =\int_{-\infty}^{+\infty}\left[\int_{-\infty}^{s} f_{R}(r;t) \mathrm{d} r\right] \cdot f_{S}(s) \mathrm{d} s
\end{aligned}
\end{equation}

Ge and Tanaka \cite{ge2000aerodynamic} elaborated the difference with various number of selected natural modes participating in the flutter by multi-mode flutter analysis, showing that the error between two modes (fundamental vertical mode and torsional mode) and multiple modes (up to 7 modes included) is less than $0.3\%$. As a result, two modes (fundamental vertical and torsional modes) are accurate to calculate the flutter critical wind speed. In this paper, only 1st-order asymmetric vertical mode and 1st-order asymmetric torsional mode are considered in the calculation of the flutter critical wind speed. In order to clarify how randomness is delivered from structural properties to the flutter critical wind speed, a linear regression model is proposed as Eq.~\eqref{eq:linear_regression_model}

\begin{equation}
\label{eq:linear_regression_model}
V_{R} = \alpha_{v1}f_{v1} + \alpha_{t1}f_{t1} + \beta_{v1}\zeta_{v1} + \beta_{t1}\zeta_{t1} + c
\end{equation}
where $\alpha_{v1}$, $\alpha_{t1}$, $\beta_{v1}$ and $\beta_{t1}$ are regression coefficients; $c$ is constant; $f_{v1}$, $\zeta_{v1}$ and $f_{t1}$, $\zeta_{t1}$ are modal frequencies and damping ratios, respectively for 1st-order vertical mode and 1st-order torsional mode. 

If the variances of $f_{i}$ and $\zeta_{i}$ ($i=v1,t1$) in Eq.~\eqref{eq:linear_regression_model} are independent (as shown in Fig.~\ref{fig:covcoeff_measurements}), the PDF $f_{R}(r)$ of the flutter critical wind speed can be directly obtained from the structural properties' PDFs
\begin{equation}
\label{eq:randomness_flow_PDF}
\makebox[\textwidth][c]{
$f_{R}(r)=\frac{1}{\left|\alpha_{v 1} \alpha_{t 1} \beta_{v 1} \beta_{t 1}\right|} \iiint_{-\infty}^{+\infty} f_{f, v 1}\left(\frac{x_{1}}{\alpha_{v 1}}\right) f_{f, t 1}\left(\frac{x_{2}-x 1}{\alpha_{t 1}}\right) f_{\zeta, v 1}\left(\frac{x_{3}-x_{2}}{\beta_{v 1}}\right) f_{\zeta, t 1}\left(\frac{r-x_{3}-c}{\beta_{t 1}}\right) \mathrm{d} x_{1} \mathrm{d} x_{2} \mathrm{d} x_{3}$
}
\end{equation} 
where $f_{f,i}(\cdot)$ $(i=v1,t1)$ denote PDFs of modal frequencies; $f_{\zeta,i}(\cdot)$ $(i=v1,t1)$ denote PDFs of damping ratios.

\section{Assessment framework} 
The evaluation of flutter probability is a well-known issue. In this paper, a linear regression model is proposed to predict the PDF of flutter critical wind speed given that the PDFs of structural properties are known. Then the probability interference method is utilized to evaluate the life-cycle flutter probability. The whole process can be summarized as the following procedure. 
\begin{enumerate}
\item Structural vibration data is collected from the health monitoring system, and Bayesian FFT modal identification is carried out to obtain structural properties, including modal frequencies and damping ratios. Futuristic evolving trends of structural properties are predicted, as well as their probability distributions.
\item Meteorological data is processed at the structure site to evaluate the PDF of annual wind speed.
\item Obtain flutter derivatives by wind tunnel test.
\item Estimate time-variant PDFs of structural properties to get the PDF of the flutter critical wind speed for each year by the proposed linear regression model, then assess long-term bridge reliability in terms of flutter probability.
\end{enumerate}

\begin{figure}[!htb]
    \centering
    \makebox[\textwidth][c]{\includegraphics[]{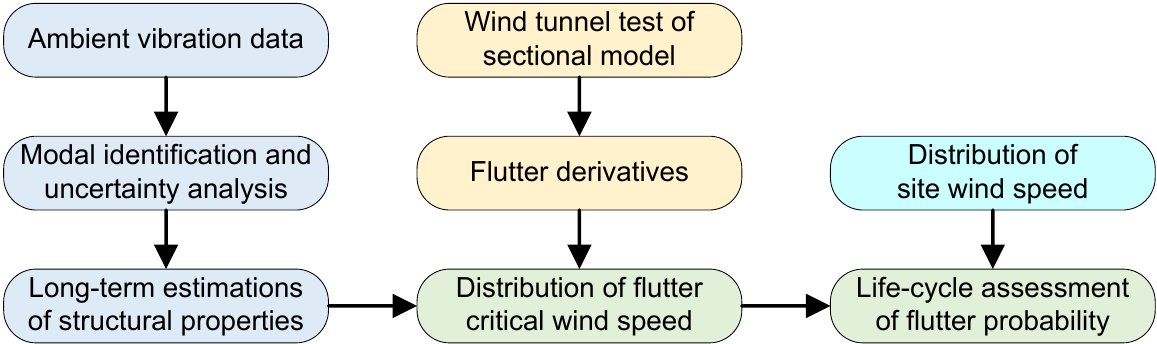}}
    \caption{Flow chart of assessment of flutter probability}
  \label{fig:framework}
\end{figure}

The whole framework is illustrated in Fig.~\ref{fig:framework}.

\section{Application example}
\label{sec:case_study}
\subsection{Description of the bridge example}
\begin{figure}[!htb]
    \centering
    \makebox[\textwidth][c]{\includegraphics[]{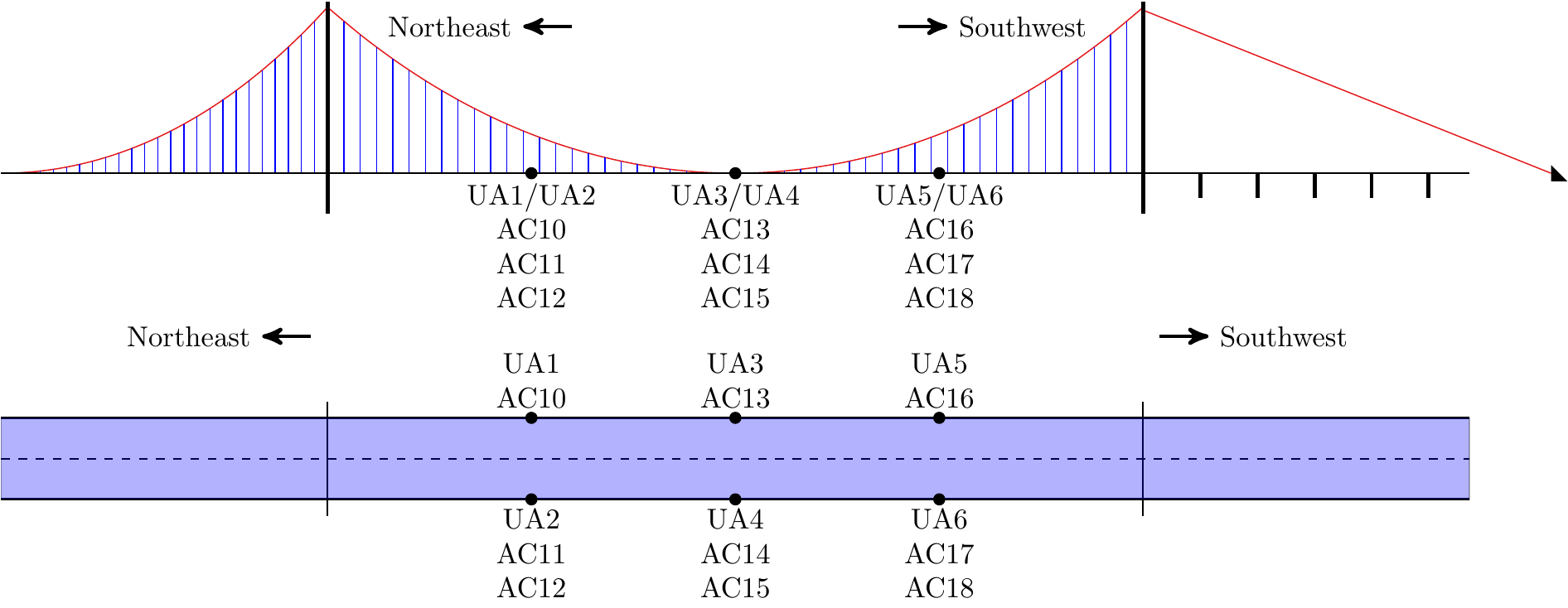}}
    \caption{Elevation of Xihoumen Bridge and layout of monitoring points}
  \label{fig:Xihoumen_Bridge}
\end{figure}

To demonstrate the feasibility of the proposed framework, this study utilizes field monitoring data from Xihoumen Bridge during 2010-2015. As shown in Fig.~\ref{fig:Xihoumen_Bridge}, Xihoumen Bridge is a suspension bridge with a 1650-meter central main span linking Jintang and Cezi islands near the East China Sea coast. In its health monitoring system, $\mathrm{UA}$1-$\mathrm{UA}$6 are ultrasonic anemometers (UAs); $\mathrm{AC}$10-$\mathrm{AC}$18 are servo accelerometers (ACs). 
The sampling rate of the UAs is 32Hz, and the measuring range is 0-65m/s. UAs can record wind speeds for 3 directions simultaneously: north, west, and upward (vertical).
The sampling rate of the ACs is 100Hz in 2010 and 2011, 50Hz in 2012-2015. ACs can record accelerations of vertical and lateral directions of the bridge cross section. 
The detailed information of UAs and ACs is presented in Tab.~\ref{tab:sensors} and Fig.~\ref{fig:Xihoumen_Bridge}.

\begin{table}[htb]
  \centering
  \caption{Definition of sensors}
  \label{tab:sensors}
  \normalsize
  \begin{tabular}{ccc}
    \toprule
    Sensor number & Sensor type & Location \\
    \midrule
    UA1/UA2&ultrasonic anemometer&main span (1/4)\\
    UA3/UA4&ultrasonic anemometer&main span (1/2)\\
    UA5/UA6&ultrasonic anemometer&main span (3/4)\\
    AC10/AC11/AC12&servo accelerometer&main span 1/4 (vertical/lateral/vertical)\\
    AC13/AC14/AC15&servo accelerometer&main span 1/2 (vertical/lateral/vertical)\\
    AC16/AC17/AC18&servo accelerometer&main span 3/4 (vertical/lateral/vertical)\\
    \bottomrule
  \end{tabular}
\end{table}

In this study, the authors utilize vertical responses at two sides of the bridge deck (i.e., $\mathrm{AC} 10$, $\mathrm{AC} 12$, $\mathrm{AC} 13$, $\mathrm{AC} 15$, $\mathrm{AC} 16$, $\mathrm{AC} 18$) because flutter usually occurs as coupled vertical and torsional motions. The acceleration histories (six years in total) are analyzed by hour. Furthermore, the authors utilize hourly-averaged wind speeds at the middle span ($\mathrm{UA} 4$) as a filter to reduce identification errors. In this study, only acceleration responses with hourly-averaged wind speeds of 2-4 m/s are adopted to avoid possible vortex-induced lock-in phenomenon (as examined by Li et al. \cite{li2011investigation}, 6-10 m/s). The reason is that when the vortex-induced vibration occurs, the bridge will be characterized by single vibration mode, which is against the assumption of fast Bayesian approach that the ambient excitation is statistically random, instead of a single fixed frequency. 2-4 m/s is an appropriate range of speeds, since it not only provides sufficient external excitations, but also avoids excessive aeroelastic effects during buffeting excitation. Moreover, in order to avoid the deviation caused by possible heavy traffic flow, only monitoring data during 0~am~-~7~am is adopted.

\subsection{Bayesian FFT modal identification results}
\label{section:identification_results}
The structural properties are significant for assessment of bridge flutter-resistance performance. The fast Bayesian approach is a useful method for modal identification of long-span bridges \cite{brownjohn2018bayesian} for the output-only system under ambient excitations \cite{yuen2003bayesian}.

\begin{figure}[!htb]
    \centering
    \makebox[\textwidth][c]{\includegraphics[]{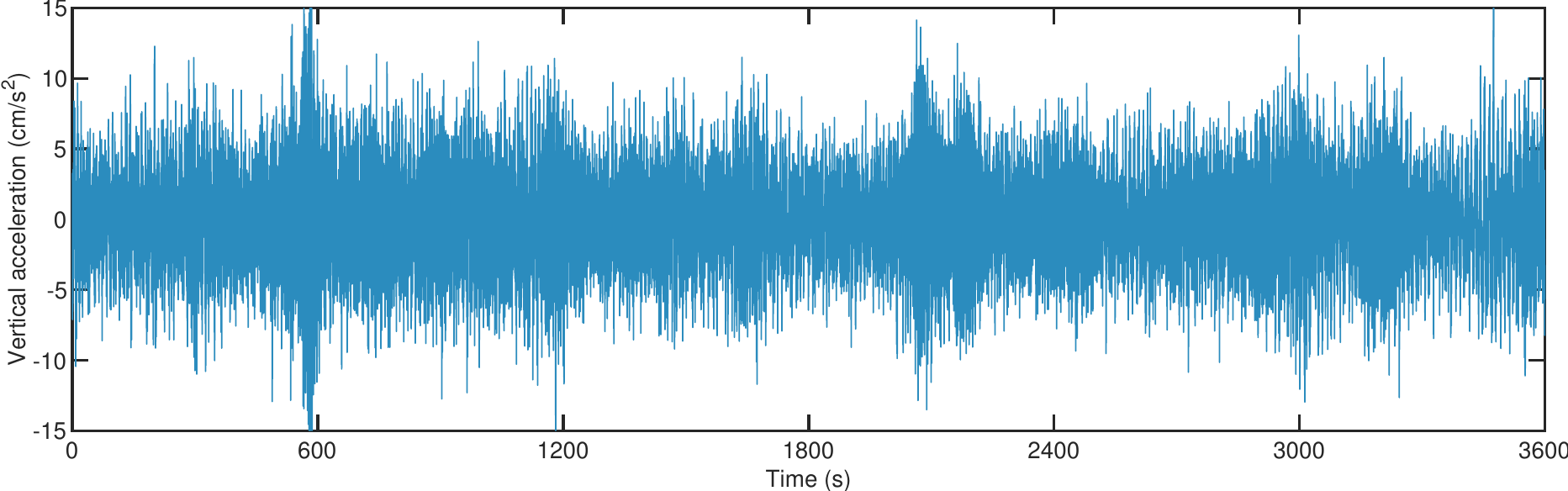}}
    \caption{Time history of one-hour vertical accelerations}
	\label{fig:acceleration_segment}
\end{figure} 

A representative segment of measured vertical acceleration is shown in Fig.~\ref{fig:acceleration_segment}. The results analyzed by SSI method \cite{li2011investigation,yang2015evaluation} and EFDD method \cite{yang2015evaluation} have been implemented by some researchers. Li et al.\cite{li2011investigation} investigated the vortex-induced vibration and structural properties (modal frequencies and damping ratios) of Xihoumen Bridge based on SSI method. Yang et al.\cite{yang2015evaluation} carried out the ambient vibration test of Xihoumen Bridge under conditions that the wind speed during the test ranged from 2m/s to 8m/s, and identify structural properties based on SSI method and EFDD method \cite{yang2015evaluation}. Tab.~\ref{tab:identified_results} shows results identified by fast Bayesian approach, as well as the results from EFDD method and SSI method for comparison. Tab.~\ref{tab:identified_results} shows that frequencies identified by different methods have a high consistency. However, there exists large discrepancies in damping ratios. The damping ratios identified by Bayesian approach are usually smaller, compared with SSI method and EFDD method. Zhang et al. \cite{zhang2019structural} also found a similar regularity that the damping ratios obtained by Bayesian approach are relatively smaller than those obtained by SSI method and EFDD method when they launched the operational modal analysis of a 250m super-tall building. This might be caused by different mechanisms while identifying damping ratios with different methods, as explained by Zhang et al. \cite{zhang2019structural}. On the other hand, different ambient conditions (e.g., temperature, humidity, wind, etc.) also make a significant difference to damping ratios \cite{li2010modal,xia2006long}. Therefore, the discrepancies shown in damping ratios are reasonable, since the field monitoring data, utilized for identification by Li et al. \cite{li2011investigation}, Yang et al. \cite{yang2015evaluation} and this study, are under fairly different ambient environments (with different temperature, humidity and so on). What's more, similar with many researchers' results  \cite{brownjohn2018bayesian,zhang2019structural}(identified by the Bayesian approach), the Bayesian coefficients of variance (COV) (=posterior standard deviation/MPV, as mentioned in Tab.~\ref{tab:identified_results}) of damping ratios are much more remarkable than those of modal frequencies. This also indicates that the changeable ambient environment influences the identified damping ratios greatly.

\begin{table}[t]
	\centering
	\caption{The measured modal frequencies and damping ratios from vertial vibration by EFDD method \cite{yang2015evaluation}, SSI method \cite{li2011investigation,yang2015evaluation} and fast Bayesian approach}
	\label{tab:identified_results}
	\scriptsize
	\begin{threeparttable}
	\makebox[\textwidth][c]{
	\begin{tabular}{cccc|ccccc|c}
		\toprule
		\multicolumn{4}{c|}{Frequency(Hz)} & \multicolumn{5}{c|}{Damping Ratio(\%)} & \multirow{3}*{Mode Shape}\\
		EFDD/SSI & SSI & Bayesian & Bayesian Frequency & EFDD & SSI & SSI & Bayesian & Bayesian Damping Ratio  \\
		Method \cite{yang2015evaluation} & Method \cite{li2011investigation}& Method & COV ($\mathrm{\sigma / MPV}$) & Method \cite{yang2015evaluation} & Method \cite{yang2015evaluation} & Method \cite{li2011investigation} & Method & COV ($\mathrm{\sigma / MPV}$) \\
		\midrule
		0.095 & 0.0953 & 0.0948 & 0.2279\% & 1.12$\sim$2.64 & 1.80$\sim$2.18& 0.57 & 0.78 & 32.48\% & 1-AS-V \\
		0.133 & 0.1328 & 0.1330 & 0.1080\% & 0.84$\sim$2.32 & 0.90$\sim$1.46& 0.52 & 0.47 & 32.57\% & 2-S-V \\
		0.183 & 0.1825 & 0.1828 & 0.0547\% & 0.18$\sim$1.02 & 0.37$\sim$0.61& 0.50 & 0.32 & 32.15\% & 2-AS-V \\
		0.229 & 0.2301 & 0.2302 & 0.0963\% & 0.21$\sim$0.59 & 0.23$\sim$0.62& 0.51 & 0.29 & 37.54\% & 1-S-T \\
		0.233 & $\backslash$ & 0.2383 & 0.0801\% & 0.25$\sim$0.41 & 0.77$\sim$0.95& $\backslash$ & 0.31 & 44.55\% & 1-AS-T \\
		0.276 & 0.2767 & 0.2767 & 0.0820\% & 0.34$\sim$1.14 & 0.43$\sim$0.83& 0.39 & 0.30 & 26.99\% & 3-AS-V \\
		\bottomrule
	\end{tabular}
	}
	\begin{tablenotes}
	\item[1] Note: S-symmetric, AS-asymmetric, V-vertical bending, T-torsion
	\end{tablenotes}
	\end{threeparttable}
\end{table}

\begin{figure}[!htb]
    \centering
    \makebox[\textwidth][c]{
    \begin{minipage}[c]{90mm}
      \centering
      \includegraphics[]{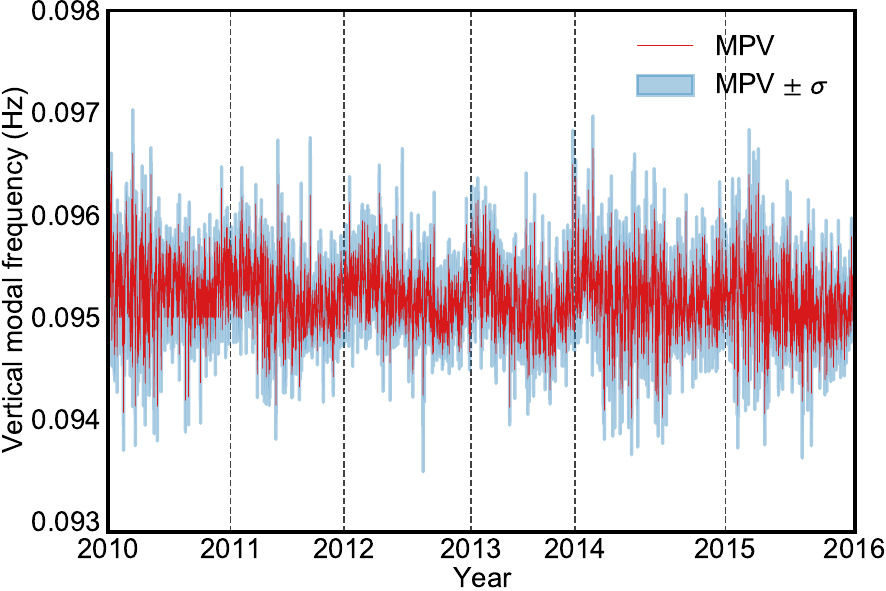}
      \caption*{(a) Vertical modal frequency}
      \label{fig:vertical_frequency}
    \end{minipage}
    \begin{minipage}[c]{90mm}
      \centering
      \includegraphics[]{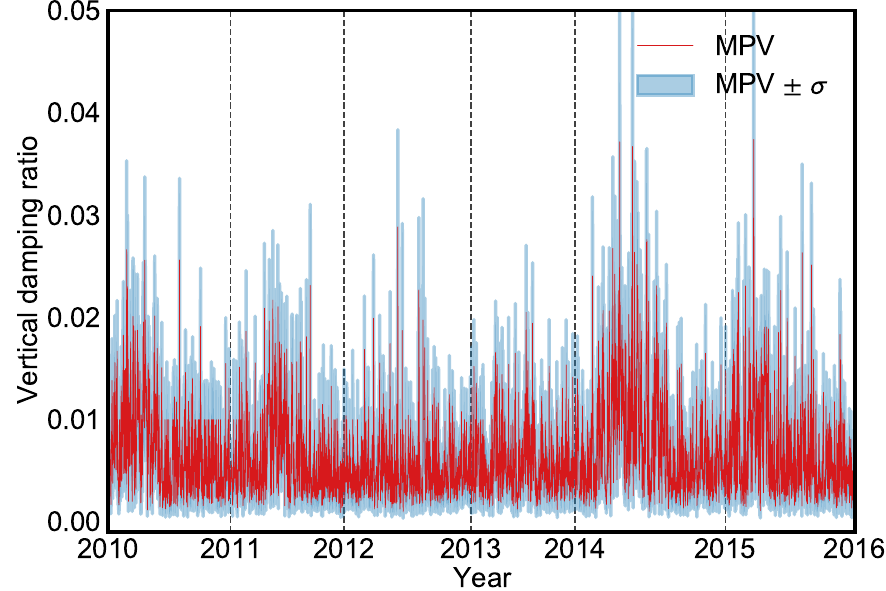}
      \caption*{(b) Vertical damping ratio}
      \label{fig:vertical_damping_ratio}
    \end{minipage}
    }
    \makebox[\textwidth][c]{
    \begin{minipage}[c]{90mm}
      \centering
      \includegraphics[]{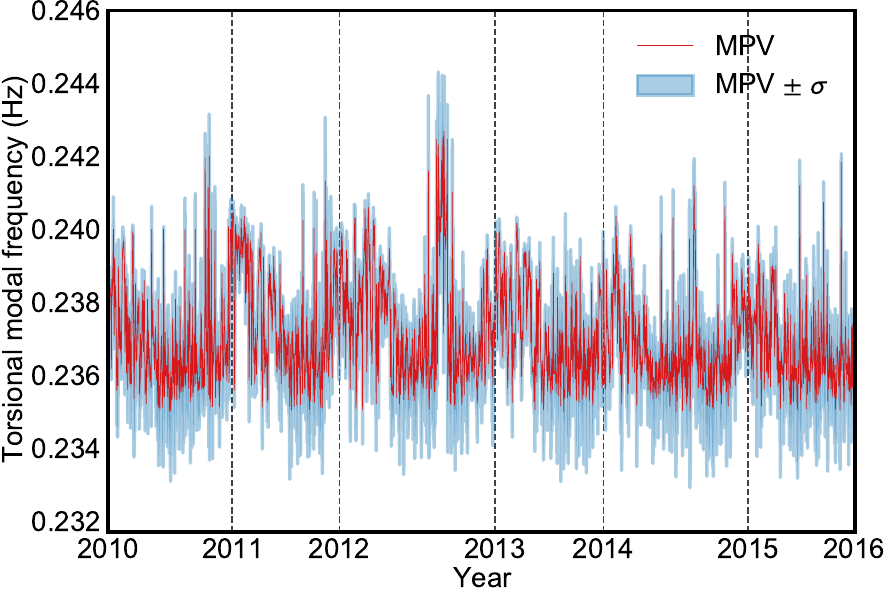}
      \caption*{(c) Torsional modal frequency}
      \label{fig:torsional_frequency}
    \end{minipage}
    \begin{minipage}[c]{90mm}
      \centering
      \includegraphics[]{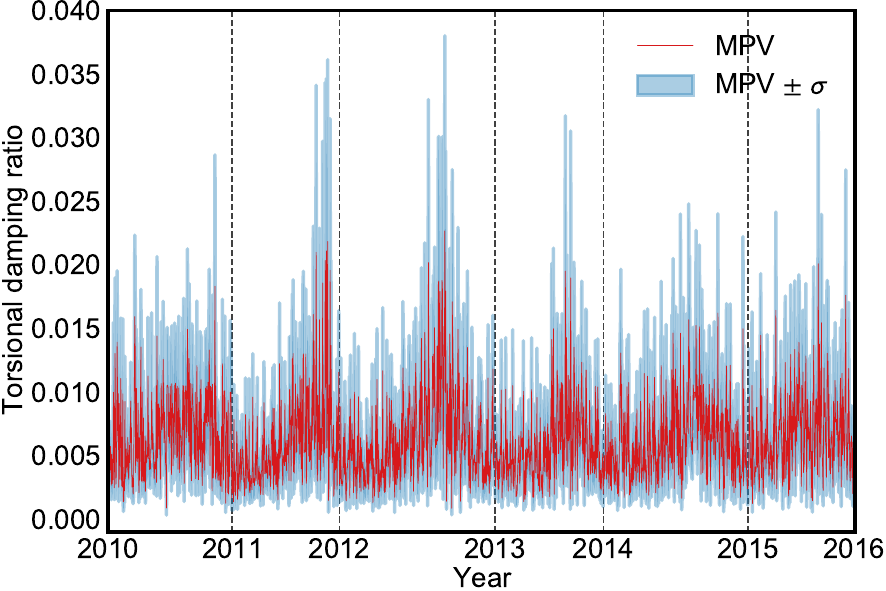}
      \caption*{(d) Torsional damping ratio}
      \label{fig:torsional_damping_ratio}
    \end{minipage}
    }
    \caption{1st-order asymmetric vertical and torsional modes: modal frequencies and damping ratios identified by fast Bayesian approach}
    \label{fig:vertical_vibration}
\end{figure}

Fig.~\ref{fig:vertical_vibration} presents the identified modal frequencies and damping ratios by fast Bayesian approach during 2010-2015, where the vibration data is hourly-segmented for analysis. 
The MPV (most probable value) is denoted by red lines. The blue area denotes the range of MPV $\pm$ posterior standard deviation (MPV $\pm$ $\sigma$), which indicates measurement errors. The variances of structural properties' MPVs show an obvious inter-seasonal fluctuating characteristic, which may be influenced by the periodic variances of environmental factors, such as temperature, humidity, etc..

For modal frequencies, they are directly associated with structural stiffness, because if the mass $M$ is assumed as constant, then a deterioration of stiffness $K$ leads to a decrease of modal frequencies ($f \propto \sqrt{K/M} \propto \sqrt{K}$). Combined with the periodic environment, modal frequencies are believed to decline generally and fluctuate inter-seasonally, as examined by Yuen et al. \cite{yuen2010ambient}. The long-term variation of damping ratios is more complicated, related with various factors \cite{li2010modal,xia2006long}. Its long-term trend could be determined by health monitoring data. Similarly with modal frequencies, damping ratios also fluctuate inter-seasonally with the periodic environment.

In order to delineate long-term trends and inter-seasonal fluctuations of structural properties more clearly, historical data from SHM system is of much significance. In Sec.~\ref{sec: analysis of structural properties}, identified long-term MPVs are utilized to fit the long-term trends and model the inter-seasonal fluctuations by probability distributions.

\subsection{Design wind speed at bridge site}
Xihoumen Bridge is a cross-sea bridge located at Zhoushan City, often invaded by typhoons \cite{zhao2019measurement}. According to the wind-resistance design specification \cite{chen2018wind}, wind velocities of all directional magnitudes at the deck height with 10-year, 50-year and 100-year return periods can be obtained, respectively 35.64 m/s, 46.48 m/s and 50.47 m/s. The Gumbel distribution \cite{chen2018wind} is utilized to fit the annual wind velocity. The probability density function (PDF) of Gumbel distribution with location parameter $\mu$ and scale parameter $\sigma$ is
\begin{equation}\label{eq:x_p}
f(s | \mu, \sigma)=\sigma^{-1} \exp \left(-\left(\frac{s-\mu}{\sigma}+\exp \left(-\left(\frac{s-\mu}{\sigma}\right)\right)\right)\right)
\end{equation}

\begin{figure}[!htb]
    \centering
    \includegraphics[]{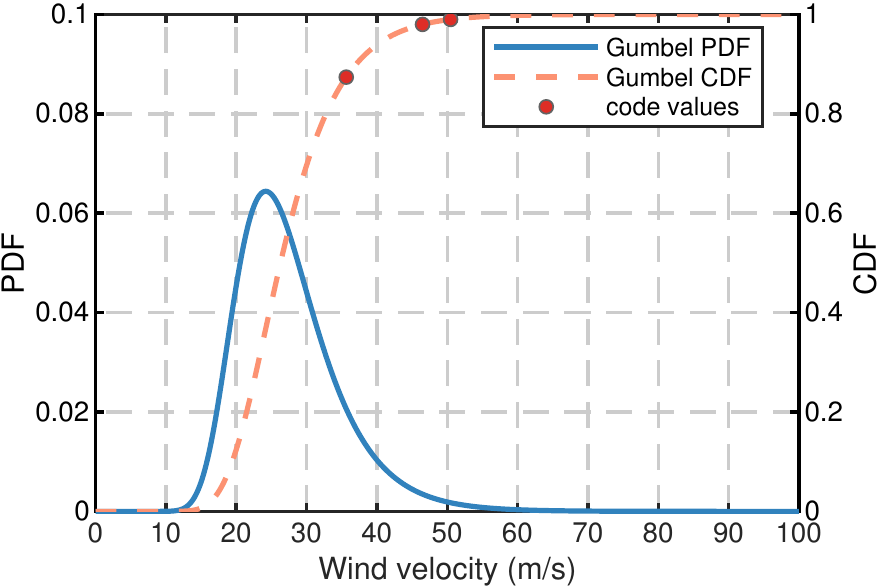}
    \caption{Probability distribution of site wind at Xihoumen Bridge}
    \label{fig:distribution_site_wind}
\end{figure}

By Eq.~\eqref{eq:x_p}, it can be deduced $\mu = 24.1973$ and $\sigma = 5.7115$. The distribution of site wind at Xihoumen Bridge is depicted in Fig.~\ref{fig:distribution_site_wind}.

\subsection{Flutter derivatives}
\label{sec: flutter derivatives}
\begin{figure}[!htb]
  \centering
  \makebox[\textwidth][c]{
    \begin{minipage}[c]{90mm}
    \centering
    \includegraphics[]{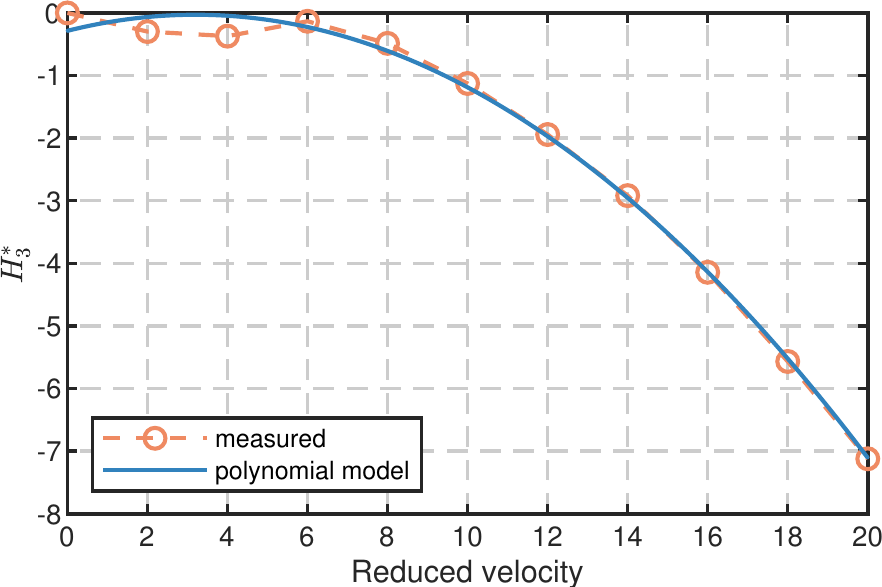}
    \caption*{(g) Flutter derivative $H_{3}^{*}$}
    \label{fig:H3}
  \end{minipage}
  \begin{minipage}[c]{90mm}
    \centering
    \includegraphics[]{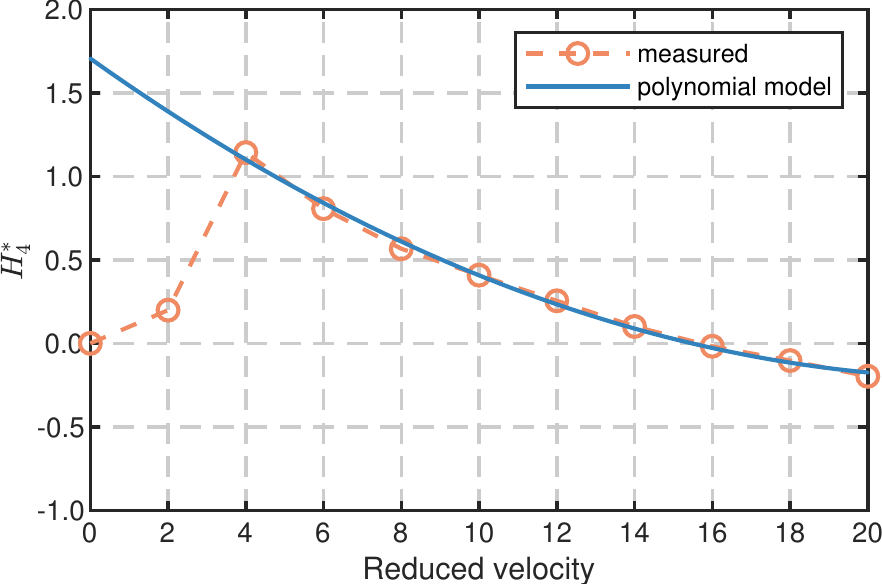}
    \caption*{(h) Flutter derivative $H_{4}^{*}$}
    \label{fig:H4}
  \end{minipage}
  }
    \caption{Flutter derivatives of Xihoumen Bridge section model}
    \label{fig:flutter_derivatives}
\end{figure}

The detailed experimental set-up information can be found in the wind tunnel test by Yang et al. \cite{yang2015evaluation}. Fig.~\ref{fig:flutter_derivatives} presents the identified flutter derivatives. All flutter derivatives are fitted by quadratic polynomial except $H_{2}^{*}$, which is fitted by quartic polynomial. As examined by Li et al. \cite{li2011investigation}, vortex-induced vibration would occur at a low wind velocity (6-10 m/s). Therefore, there will exist aberrant values at certain low reduced velocities in Fig.~\ref{fig:flutter_derivatives}(d), Fig.~\ref{fig:flutter_derivatives}(e) and Fig.~\ref{fig:flutter_derivatives}(h), where the abnormal points are excluded in order to improve the fitting accuracy. It is reasonable because in this application example, the reduced velocity will range from 8 to 13 when the bridge flutter occurs.

\begin{figure}[!htb]
    \centering
    \makebox[\textwidth][c]{
    \begin{minipage}[c]{90mm}
    \centering
    \includegraphics[]{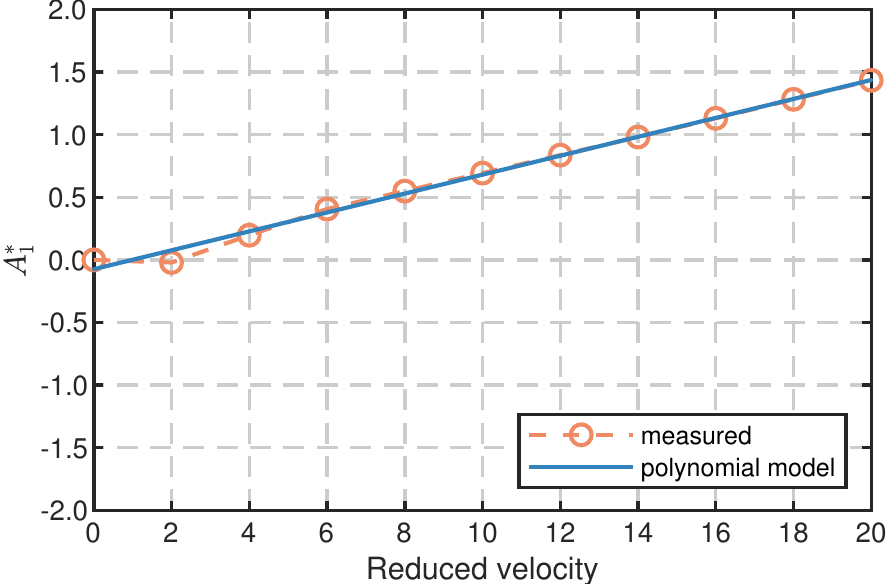}
    \caption*{(a) Flutter derivative $A_{1}^{*}$}
    \label{fig:A1}
    \end{minipage}
    \begin{minipage}[c]{90mm}
    \centering
    \includegraphics[]{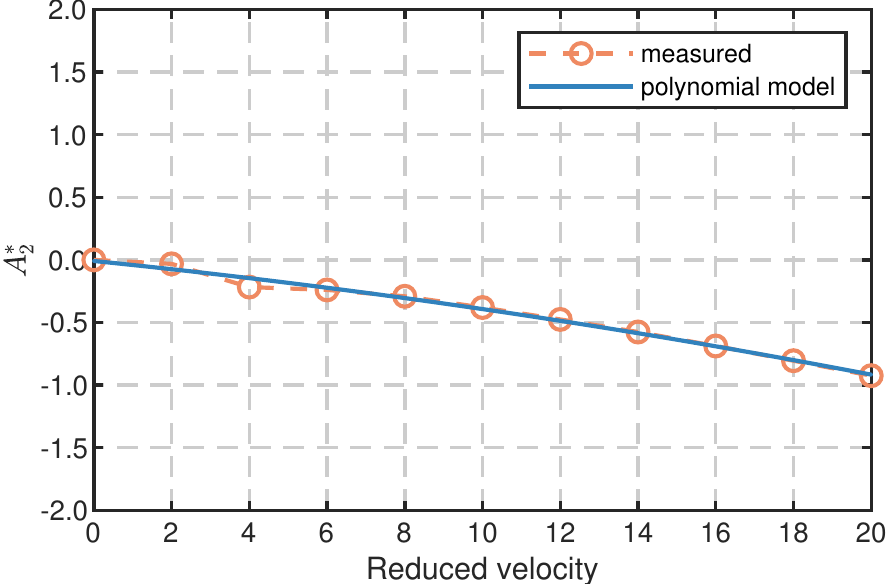}
    \caption*{(b) Flutter derivative $A_{2}^{*}$}
    \label{fig:A2}   
    \end{minipage}
    }
\end{figure}


\begin{figure}[!htb]
  \centering
  \makebox[\textwidth][c]{
  \begin{minipage}[c]{90mm}
      \centering
      \includegraphics[]{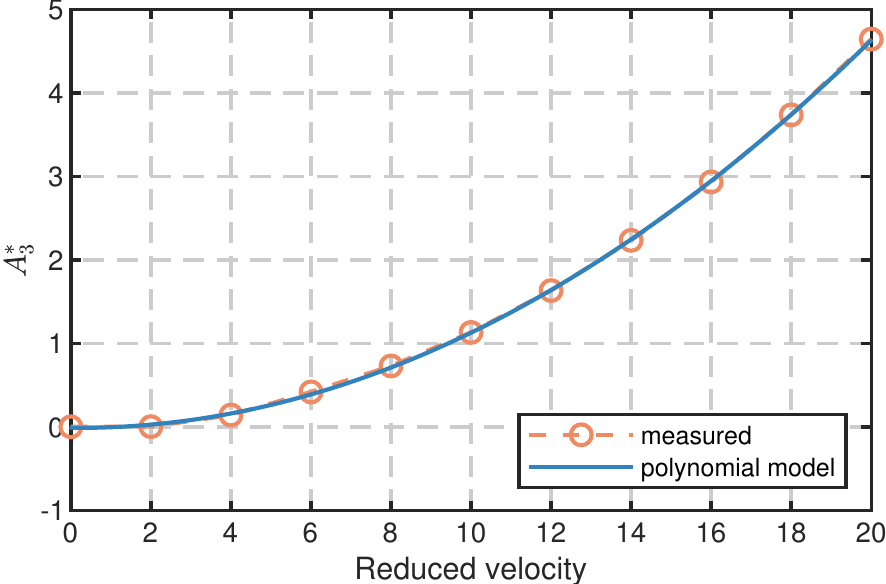}
      \caption*{(c) Flutter derivative $A_{3}^{*}$}
      \label{fig:A3}
  \end{minipage}
  \begin{minipage}[c]{90mm}
    \centering
    \includegraphics[]{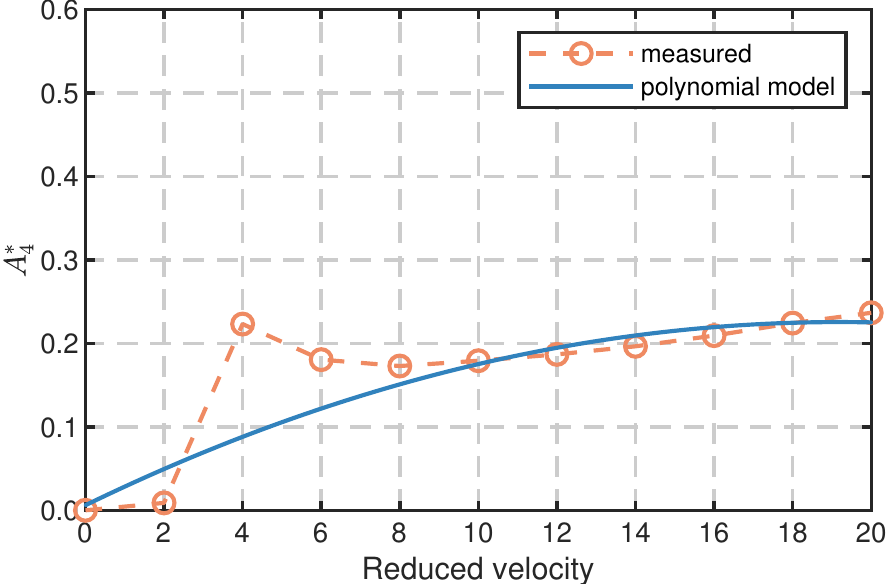}
    \caption*{(d) Flutter derivative $A_{4}^{*}$}
    \label{fig:A4}
  \end{minipage}
  }
\end{figure}

\begin{figure}[!htb]
  \centering
  \makebox[\textwidth][c]{
  \begin{minipage}[c]{90mm}
    \centering
    \includegraphics[]{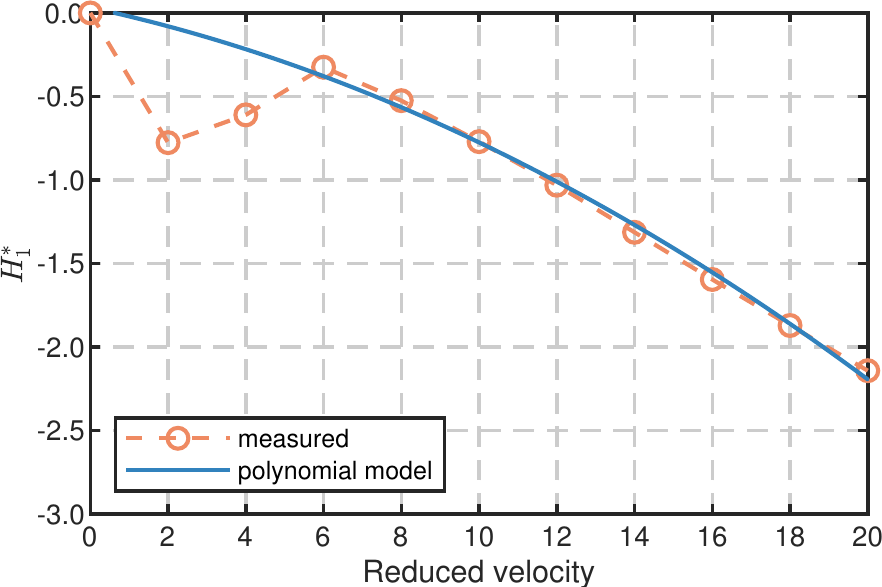}
    \caption*{(e) Flutter derivative $H_{1}^{*}$}
    \label{fig:H1}
  \end{minipage}
    \begin{minipage}[c]{90mm}
      \centering
    \includegraphics[]{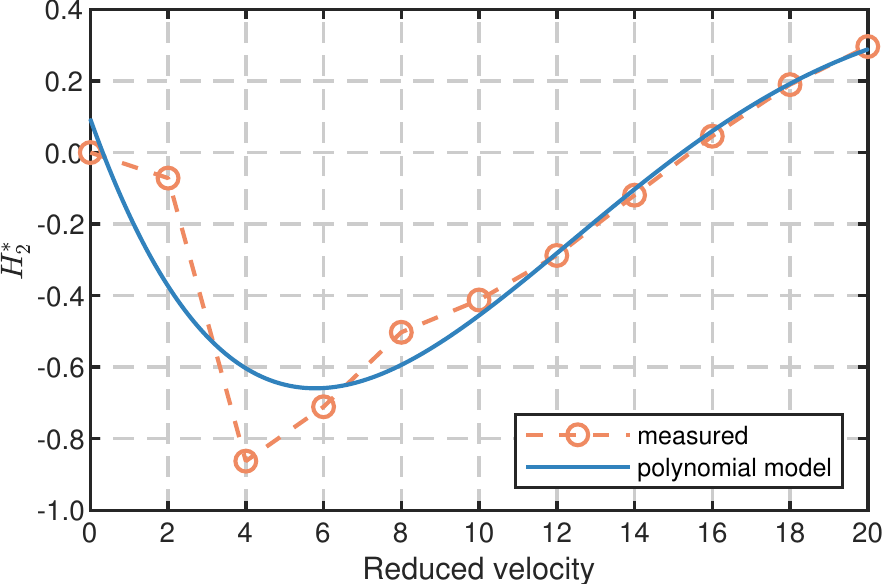}
    \caption*{(f) Flutter derivative $H_{2}^{*}$}
    \label{fig:H2}   
    \end{minipage}
    }
\end{figure}

\section{Dynamic properties}
\label{sec: analysis of structural properties}
\subsection{Modal frequencies}
\label{sec:structural_properties_prediction}
\begin{figure}[!htb]
    \centering
    \makebox[\textwidth][c]{
    \begin{minipage}[c]{90mm}
      \centering
      \includegraphics[]{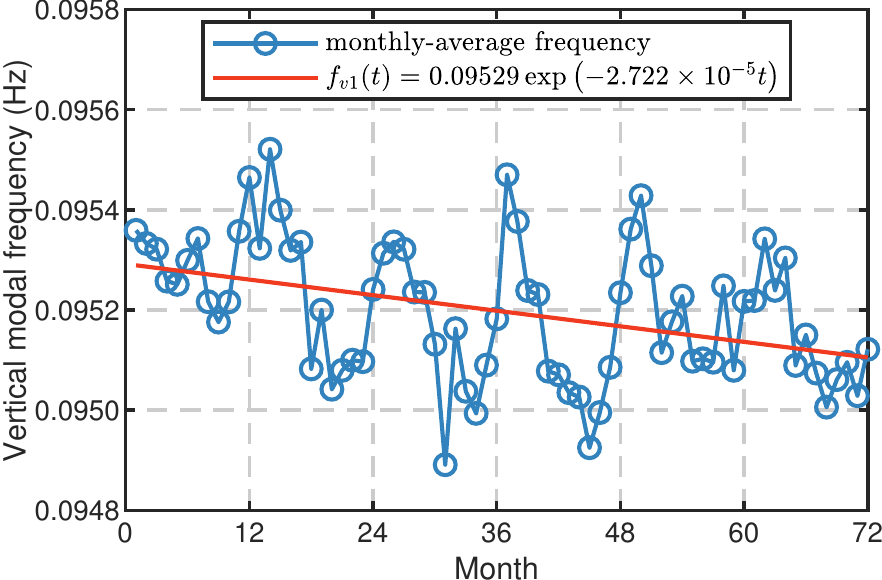}
      \caption*{(a) Monthly-averaged vertical modal frequency}
    \label{fig:vertical_frequency_fitting}
      \end{minipage}
    \begin{minipage}[c]{90mm}
      \centering
      \includegraphics[]{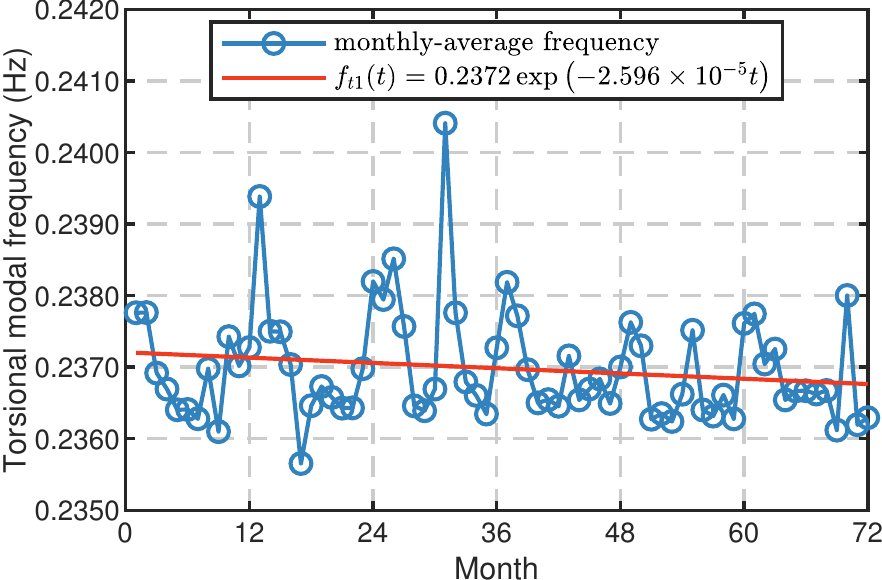}
      \caption*{(b) Monthly-averaged torsional modal frequency}
    \label{fig:torsional_frequency_fitting}
    \end{minipage}
    }
    \caption{1st-order asymmetric vertical and torsional modes: monthly-averaged modal frequencies and corresponding fitting curves}
    \label{fig:monthly_average_modal_frequencies_predictions}
\end{figure}

In this section, a solution is proposed to describe the time-variant structural properties, where the structural deterioration due to aging is modeled by a time-variant function and the fluctuation due to inter-seasonal environmental effects is described by probability distributions. In order to eliminate the unavoidable data recoding gaps of SHM, the MPVs of modal frequencies are averaged monthly in Fig.~\ref{fig:monthly_average_modal_frequencies_predictions}. The time-variant deterioration of modal frequencies is a function of time, unit of which is month. Meanwhile, probability distributions of the fluctuation can be obtained from regression residuals of the time-variant functions.

\subsubsection{Time-variant deterioration functions}
\label{sec:deterioration_frequencies}
Fig.~\ref{fig:monthly_average_modal_frequencies_predictions} presents the changing trends and variations of the monthly-averaged MPVs of modal frequencies. It shows obviously that the modal frequencies have been decreasing generally with time due to aging and fluctuating inter-seasonally. To quantitatively describe the time-variant deterioration, an exponential function $f(t) = a\exp(bt)$ is utilized to perform the regression analysis, where the unit of $t$ is month. The time-variant deterioration functions fitted in Fig.~\ref{fig:monthly_average_modal_frequencies_predictions} are based on historical data (2010-2015) of the SHM system for Xihoumen Bridge. According to the deterioration functions, the predicted vertical modal frequency will decline from 0.09529 to 0.09223 in 100 years later, about 3.2\%. Similarly, the predicted torsional modal frequency will decline from 0.2372 to 0.2299, about 3.1\%.

\subsubsection{Probability distributions of inter-seasonal fluctuations}
\label{sec:inter_seasonal_frequencies}
As shown in Fig.~\ref{fig:monthly_average_modal_frequencies_predictions}, the MPVs fluctuate inter-seasonally. The probability distributions of inter-seasonal fluctuations can be fitted by 72-month regression residuals. Normal distribution and generalized extreme value (GEV) distribution are utilized to describe these fluctuations. The PDF of Normal distribution with location parameter $\mu$ and scale parameter $\sigma$ is
\begin{equation}
f(x | \mu, \sigma)=\frac{1}{\sqrt{2 \pi} \sigma} \exp \left(-\frac{(x-\mu)^{2}}{2 \sigma^{2}}\right)
\end{equation}

The PDF of GEV distribution with location parameter $\mu$, scale parameter $\sigma$, and shape parameter $k$ ($k \neq 0$) is
\begin{equation}\label{eq:gev_distribution}
f(x | k, \mu, \sigma)=\left(\frac{1}{\sigma}\right) \exp \left(-\left(1+k \frac{(x-\mu)}{\sigma}\right)^{-\frac{1}{k}}\right)\left(1+k \frac{(x-\mu)}{\sigma}\right)^{-1-\frac{1}{k}}
\end{equation}

\begin{figure}[!htb]
    \centering
    \makebox[\textwidth][c]{
    \begin{minipage}[c]{90mm}
      \centering
      \addtocounter{subfigure}{+2}
      \includegraphics[]{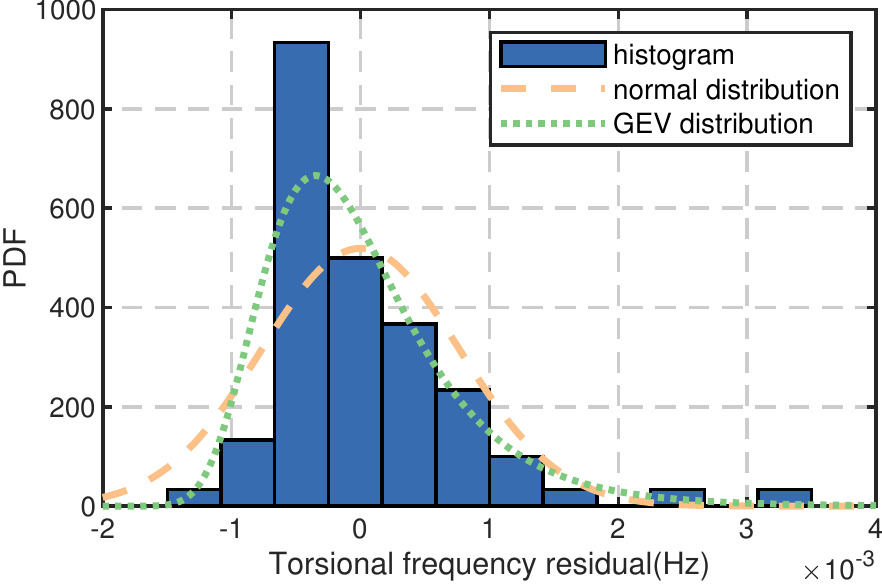}
      \caption*{(c) PDF of the torsional residual}
      \label{fig:torsional_temperaturecausederror_frequency}
    \end{minipage}
    \begin{minipage}[c]{90mm}
      \centering
      \includegraphics[]{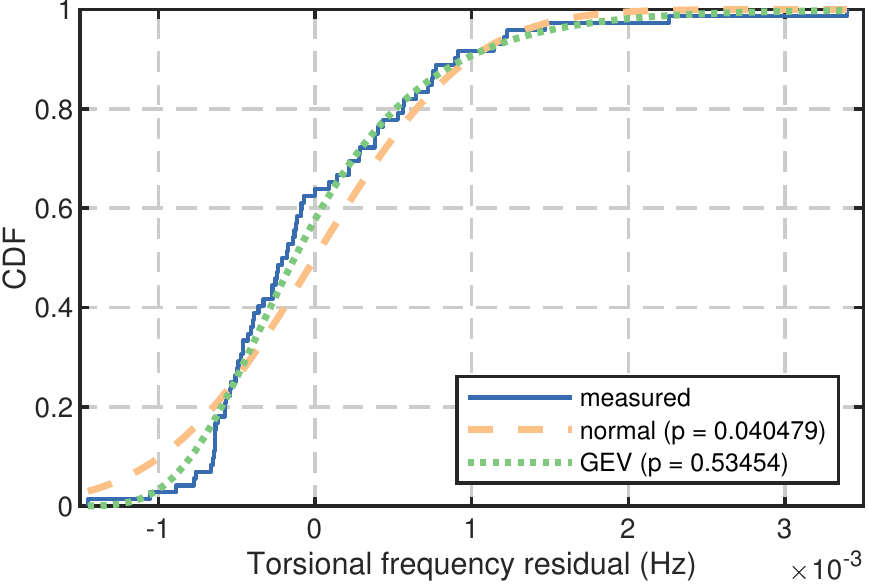}
      \caption*{(d) CDF of the torsional residual}
      \label{fig:torsional_cdf_frequency}   
    \end{minipage}
    }
	 \caption{1st-order asymmetric vertical and torsional modes: PDF and CDF of inter-seasonal fluctuations of modal frequencies, fitted by 72-month regression residuals}
    \label{fig:vertical_frequency_residual_distribution}
\end{figure}

Especially, Kolmogorov-Smirnov test (ks-test) is utilized to verify the goodness of the fitted probability distributions. The significance $p$-value is set as 0.05. Fig.~\ref{fig:vertical_frequency_residual_distribution} illustrates the fitted Normal distribution curves, GEV distribution curves and their corresponding ks-test values. For the 1st-order asymmetric vertical mode, Normal distribution and GEV distribution can both meet the requirement of ks-test with $p>0.05$. For the 1st-order asymmetric torsional mode, only GEV distribution is qualified with $p=0.53>0.05$, and Normal distribution is rejected with $p=0.04<0.05$. In this study, GEV distributions are adopted for both modes.

\begin{figure}[!htb]
    \centering
    \makebox[\textwidth][c]{
    \begin{minipage}[c]{90mm}
    	\centering
    	\includegraphics[]{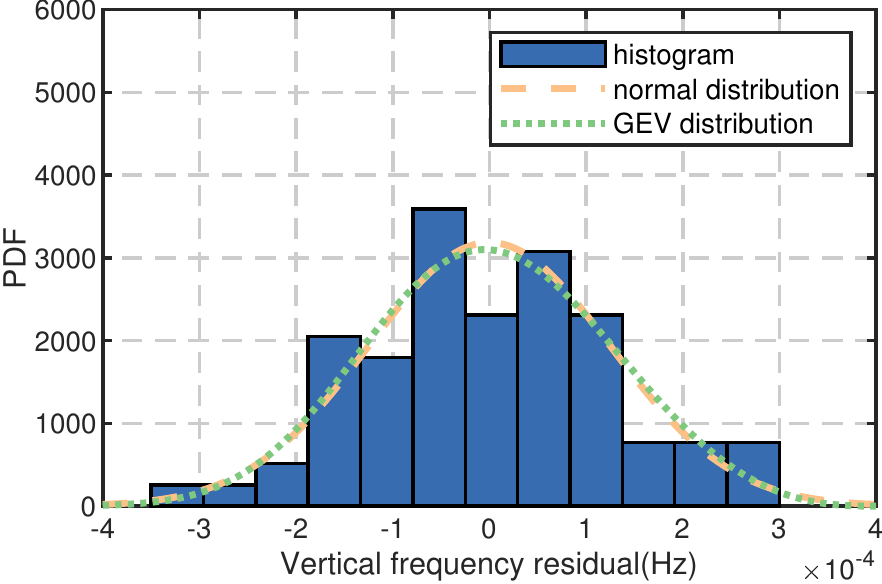}
    	\caption*{(a) PDF of the vertical residual}
    	\label{fig:vertical_temperaturecausederror_frequency}
    \end{minipage}
    \begin{minipage}[c]{90mm}
    	\centering
    	\includegraphics[]{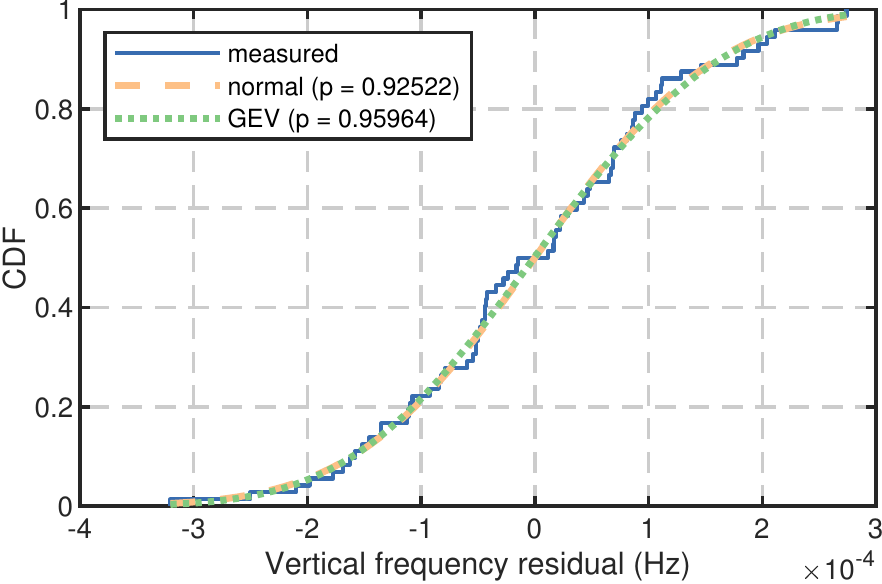}
    	\caption*{(b) CDF of the vertical residual}
    	\label{fig:vertical_cdf_frequency}   
    \end{minipage}
    }
\end{figure}

\subsubsection{Long-term evolutions of modal frequencies}
\begin{figure}[!htb]
    \centering
    \makebox[\textwidth][c]{
    \begin{minipage}[c]{90mm}
      \centering
      \includegraphics[]{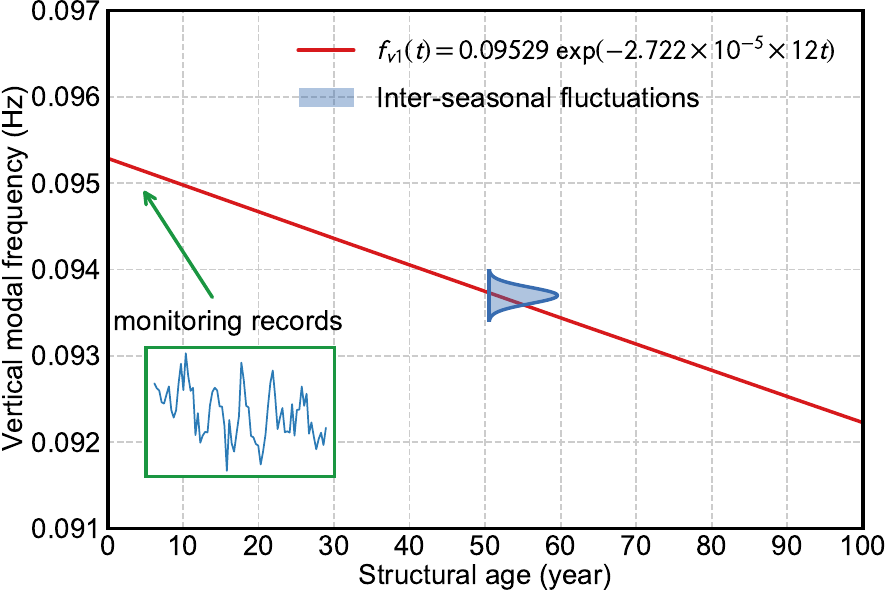}
      \caption*{(a) Evolution of 1st-order vertical modal frequency in 100-year structural age}
      \label{fig:ver_frequency_plot}
    \end{minipage}
    \begin{minipage}[c]{90mm}
      \centering
      \includegraphics[]{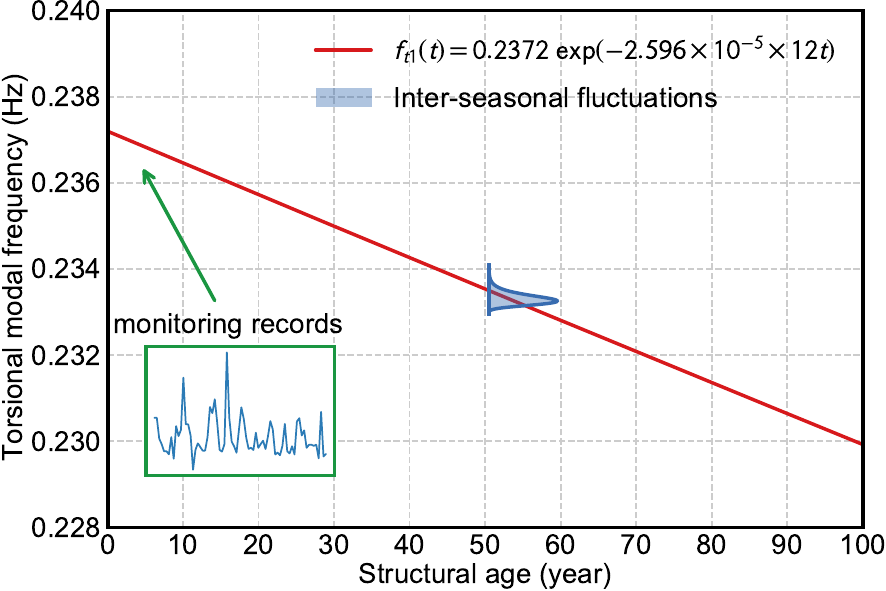}
      \caption*{(b) Evolution of 1st-order torsional modal frequency in 100-year structural age}
      \label{fig:tor_frequency_plot}   
    \end{minipage}
    }
	\caption{1st-order asymmetric vertical and torsional modes: prediction of modal frequencies in futuristic 100-year structural age, based on first 6-year monitoring records}
    \label{fig:ver_tor_plot}
\end{figure}

In Fig.~\ref{fig:ver_tor_plot}, futuristic 100-year modal frequencies are predicted, based on first 6-year monitoring records. The unit of the $x$ axis here is year rather than month, which leads to a modification of deterioration functions in terms of time $t$, as $f_{v1}(t)$ and $f_{t1}(t)$ in Fig.~\ref{fig:ver_tor_plot} show. In the long term, the modal frequencies can be described as the combination of a deterministic value (obtained by $f_{v1}(t)$ and $f_{t1}(t)$) and a random variable (blue areas of the inter-seasonal fluctuations), as illustrated in Fig.~\ref{fig:ver_tor_plot}. In this study, the probability distribution of the inter-seasonal fluctuations are time-invariant inferred from Fig.~\ref{fig:monthly_average_modal_frequencies_predictions}, which prefigures that only mean values of modal frequencies' PDFs are changing with time, but standard variances of modal frequencies' PDFs are time-invariant. Modal frequencies' PDFs will not vary sharply from January to December of the same year. As a result in Sec.~\ref{sec:distribution_flutter_limit}, the modal frequencies' PDFs in June of each year are chosen as the representative probability distributions to calculate the PDF of the flutter critical wind speed in that year.

\subsection{Damping ratios}
\label{sec:damping_ratios_probability_distribution}
\begin{figure}[!htb]
    \centering
    \makebox[\textwidth][c]{
    \begin{minipage}[c]{90mm}
      \centering
    \includegraphics[]{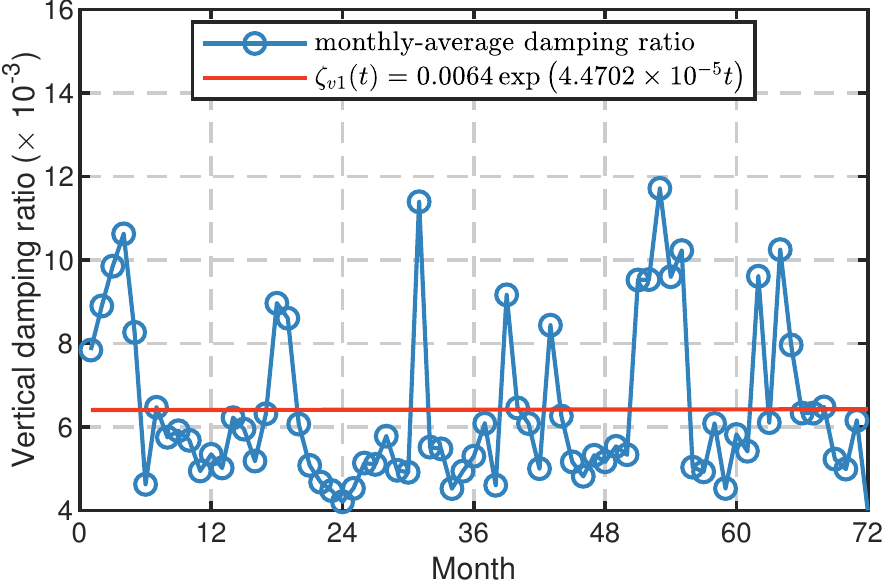}
    \caption*{(a) Monthly-averaged vertical damping ratio}
    \label{fig:vertical_monthlyaverage_damping_ratio}
    \end{minipage}
    \begin{minipage}[c]{90mm}
      \centering
    \includegraphics[]{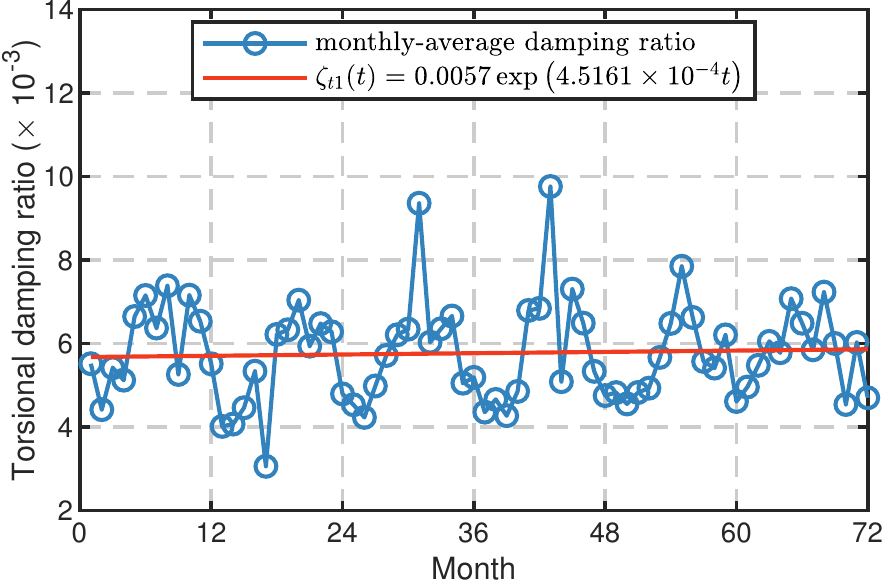}
    \caption*{(b) Monthly-averaged torsional damping ratio}
    \label{fig:torsional_monthlyaverage_damping_ratio}
    \end{minipage}
    }
    \caption{1st-order asymmetric vertical and torsional modes: monthly-averaged damping ratios and corresponding fitting curves}
    \label{fig:monthly_average_damping_ratio}
\end{figure}

MPVs of damping ratios are also averaged monthly, as presented in Fig.~\ref{fig:monthly_average_damping_ratio}. Compared with modal frequencies, the long-term deterioration of damping ratios is ambiguous, but the inter-seasonal fluctuation is much more remarkable. Therefore, the long-term changing trend of damping ratios is negligible. Thus, only the inter-seasonal fluctuations are considered for damping ratios. Noticeably, with futuristic monitoring data, the deterioration effect of damping ratios might be more clear and non-negligible, which remains to be investigated further. 

Lognormal distribution, Gamma distribution and GEV distribution are utilized to model the inter-seasonal fluctuations of damping ratios. 

The PDF of Lognormal distribution is
\begin{equation}f(x | \mu, \sigma)=\frac{1}{x \sigma \sqrt{2 \pi}} \exp \left(-\frac{(\ln x-\mu)^{2}}{2 \sigma^{2}}\right)
\end{equation}
where $\mu$ and $\sigma$ are the location parameter and scale parameter, respectively.

The PDF of Gamma distribution is 
\begin{equation}
f(x | a, b)=\frac{1}{b^{a} \Gamma(a)} x^{a-1} e^{\frac{-x}{b}}
\end{equation}
where $\Gamma$($\cdot$) is the Gamma function, $a$ is a shape parameter, $b$ is a scale parameter.

\begin{figure}[!t]
 	\centering
    \makebox[\textwidth][c]{
    \begin{minipage}[c]{90mm}
    \centering
    \includegraphics[]{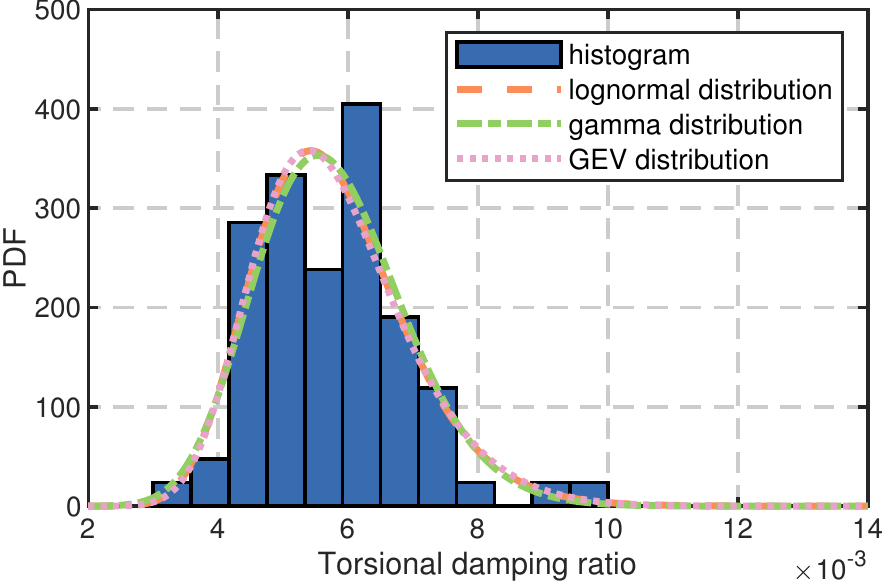}
    \caption*{(c) PDF of the torsional damping ratio}
    \label{fig:torsional_monthlyaverage_damping_ratio_distribution}
    \end{minipage}
    \begin{minipage}[c]{90mm}
      \centering
      \includegraphics[]{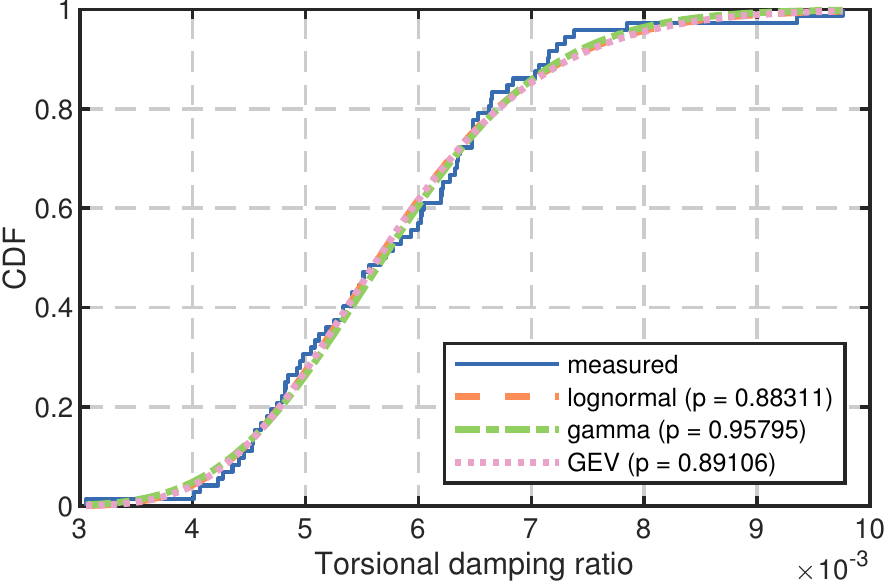}
      \caption*{(d) CDF of the torsioal damping ratio}
      \label{fig:torsional_cdf_damping_ratio}   
    \end{minipage}
    }
	\caption{1st-order asymmetric vertical and torsional modes: PDF and CDF of monthly-averaged damping ratios}
    \label{fig:vertical_damping_ratio_distribution_cdf}
\end{figure}

Fig.~\ref{fig:vertical_damping_ratio_distribution_cdf} illustrates the fitted distribution curves and their corresponding ks-test results. 
For 1st-order vertical mode, GEV distribution fits well with $p=0.57>0.05$, yet Lognormal distribution and Gamma distribution are both rejected with $p<0.05$. For 1st-order torsional mode, all three distributions fit well with $p>0.05$. In this study, the distributions with largest $p$ values are employed. Hence, GEV distribution and Gamma distribution are adopted to fit the inter-seasonal fluctuations of 1st-order vertical and torsional modes, respectively.

\begin{figure}[!b]
    \centering
    \makebox[\textwidth][c]{
    \begin{minipage}[c]{90mm}
		\centering
		\includegraphics[]{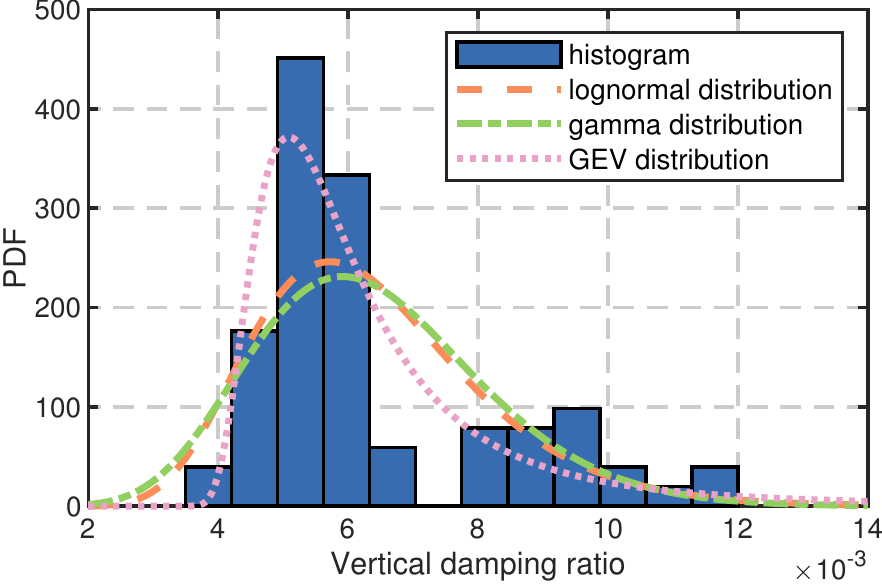}
		\caption*{(a) PDF of the vertical damping ratio}
		\label{fig:vertical_monthlyaverage_damping_ratio_distribution}
	\end{minipage}
    \begin{minipage}[c]{90mm}
    	\centering
    	\includegraphics[]{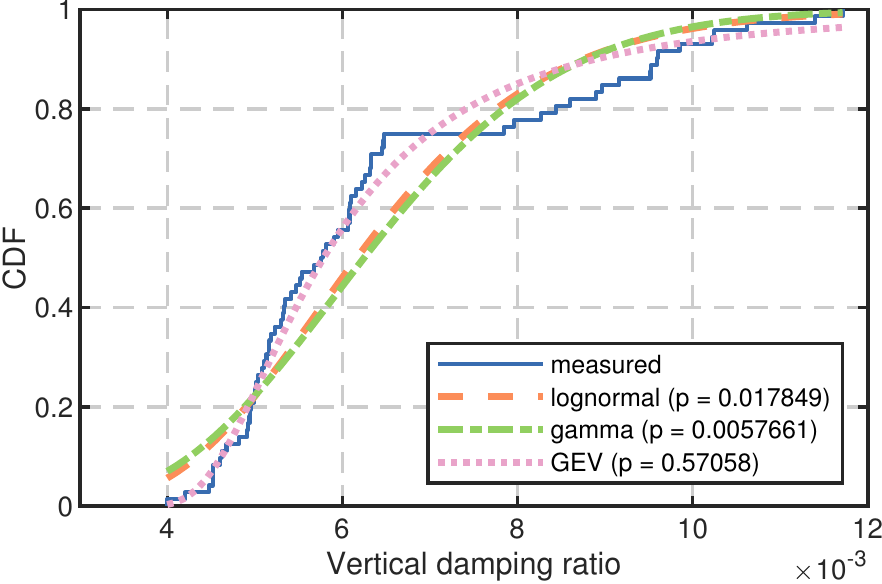}
    	\caption*{(b) CDF of the vertical damping ratio}
    	\label{fig:vertical_cdf_damping_ratio}   
    \end{minipage}
    }
 \end{figure}
 
\subsection{Correlation of modal frequencies and damping ratios}
\begin{figure}[!htb]
    \centering
    \makebox[\textwidth][c]{
    \begin{minipage}[c]{90mm}
    \centering
    \includegraphics[]{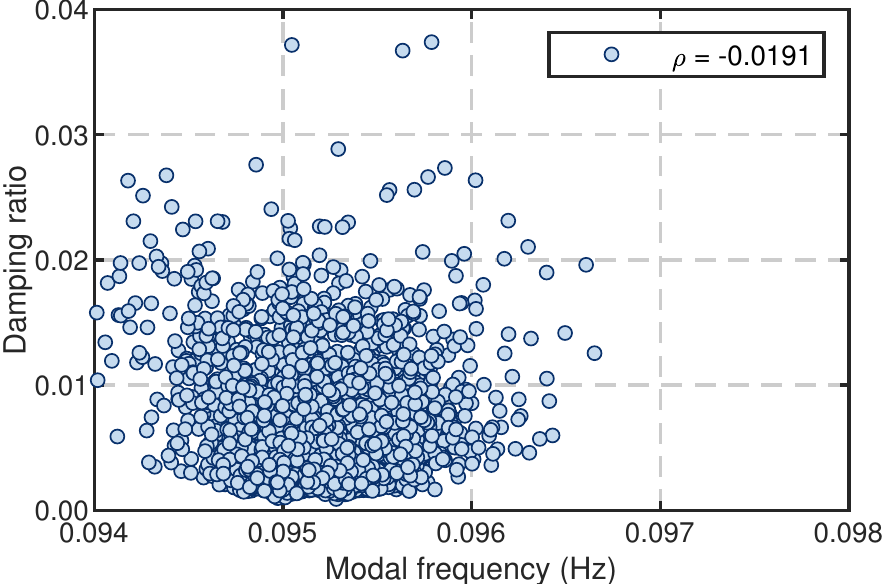}
    \caption*{(a) Vertical modal frequency and damping ratio}
    \label{fig:covcoeff_vertical}
    \end{minipage}
    \begin{minipage}[c]{90mm}
      \centering
      \includegraphics[]{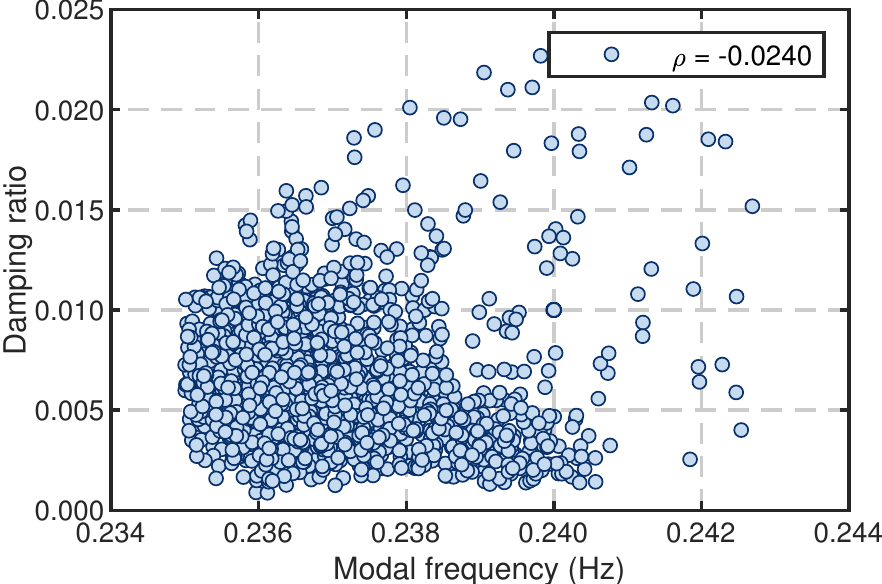}
      \caption*{(b) Torsional modal frequency and damping ratio}
      \label{fig:covcoeff_torsional}   
    \end{minipage}
    }
    \caption{Correlations for modal frequencies and damping ratios of 1st-order asymmetric modes, based on first 6-year monitoring data}
    \label{fig:covcoeff_measurements}
\end{figure}

In Fig.~\ref{fig:covcoeff_measurements}, the identified modal frequencies and damping ratios of 1st-order asymmetric vertical and torsional modes are plotted to verify their correlations, based on 6-year monitoring data. It shows that the correlation coefficients are -0.0191 and -0.0240, respectively for the vertical mode and torsional mode. Their correlations are not statistically significant. Thus for the same mode, the variance of the modal frequency is independent with the variance of the damping ratio, which leads to the assumption in Eq.~\eqref{eq:randomness_flow_PDF} that the variances of $f_{i}$ and $\zeta_{i}$ ($i=v1,t1$) are independent.

\section{Numerical simulation of life-cycle flutter probability}
\label{sec:numerical simulation}
\subsection{Distribution of flutter critical wind speed}
\label{sec:distribution_flutter_limit}
\begin{figure}[!htb]
    \centering
    \makebox[\textwidth][c]{\includegraphics[]{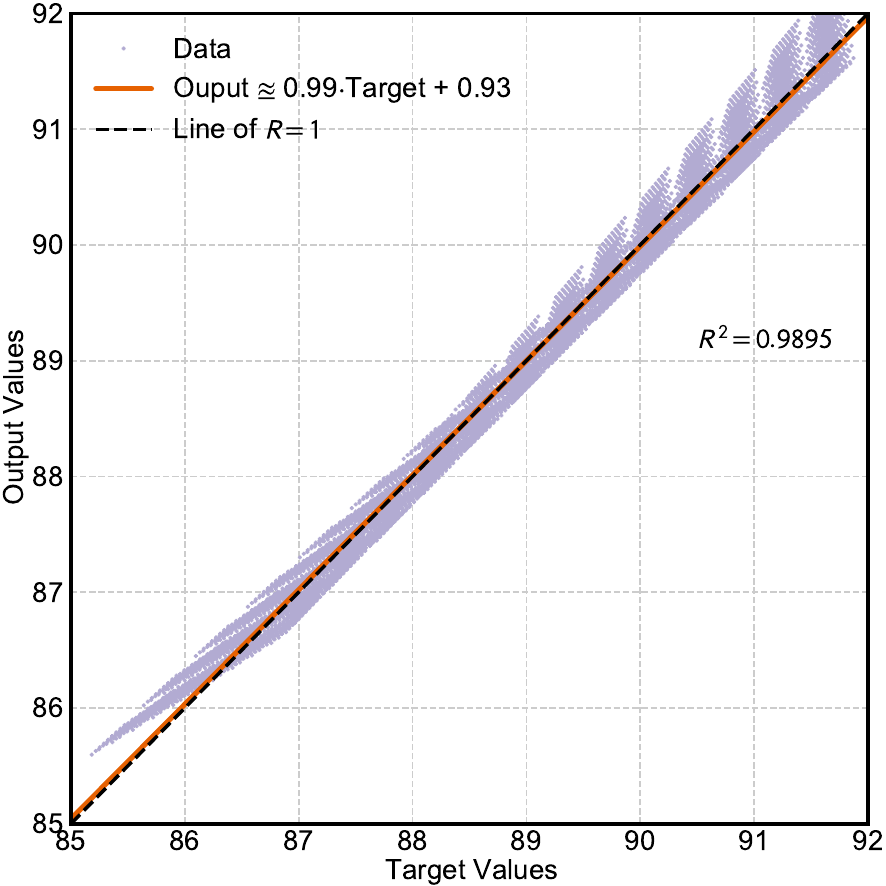}}
    \caption{Performance of the proposed linear regression model between the flutter critical wind speed and 1st-order asymmetric modal properties}
    \label{fig:linear_regression_performance}
\end{figure}

The flutter critical wind speed is calculated by the method suggested in Sec.~\ref{sec:flutter critical wind speed}. In order to incorporate the entire possible life-cycle structural properties, ranges of $f_{v1}$, $f_{t1}$, $\zeta_{v1}$ and $\zeta_{t1}$ are set as (0.090,0.100), (0.226,0.238), (0.004, 0.012) and (0.003,0.010), respectively. Then the regression coefficients $\alpha_{v1}$, $\alpha_{t1}$, $\beta_{v1}$, $\beta_{t1}$ and $c$ in Eq.~\eqref{eq:randomness_flow_PDF} can be obtained. The performance of the proposed linear regression model is good with $R^{2} \approx 0.99$, as illustrated in Fig.~\ref{fig:linear_regression_performance}.

\begin{figure}[!htb]
    \centering
    \makebox[\textwidth][c]{\includegraphics[]{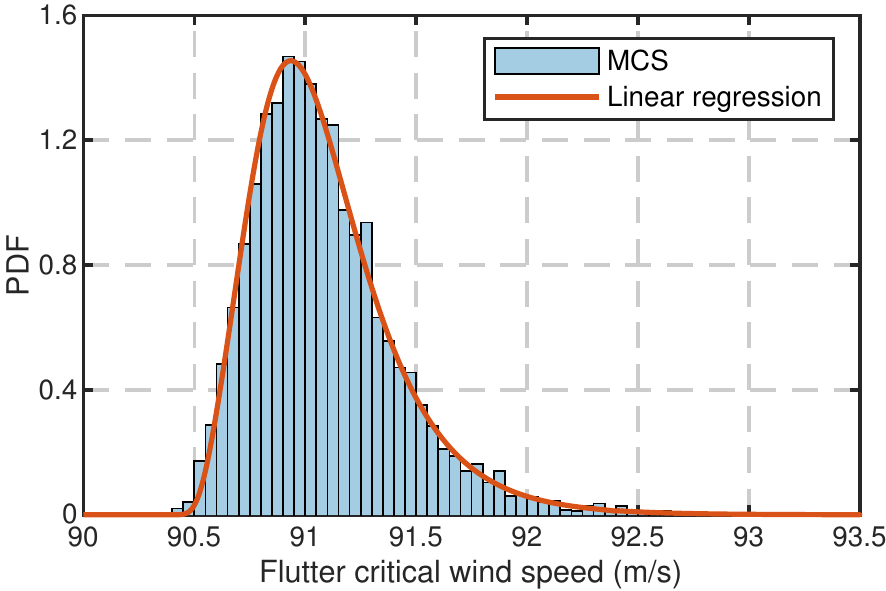}}
    \caption{Validation of the proposed linear regression model with MCS: distribution of the flutter critical wind speed at the beginning stage}
    \label{fig:Validation_MCS_linear_regression}
\end{figure}

Fig.~\ref{fig:Validation_MCS_linear_regression} presents the distribution of the flutter critical wind speed at the beginning stage (structural age = 0), where the Eq.~\eqref{eq:randomness_flow_PDF} is validated with the Monte-Carlo Simulation (MCS). 

\begin{figure}[!htb]
    \centering
    \makebox[\textwidth][c]{\includegraphics[]{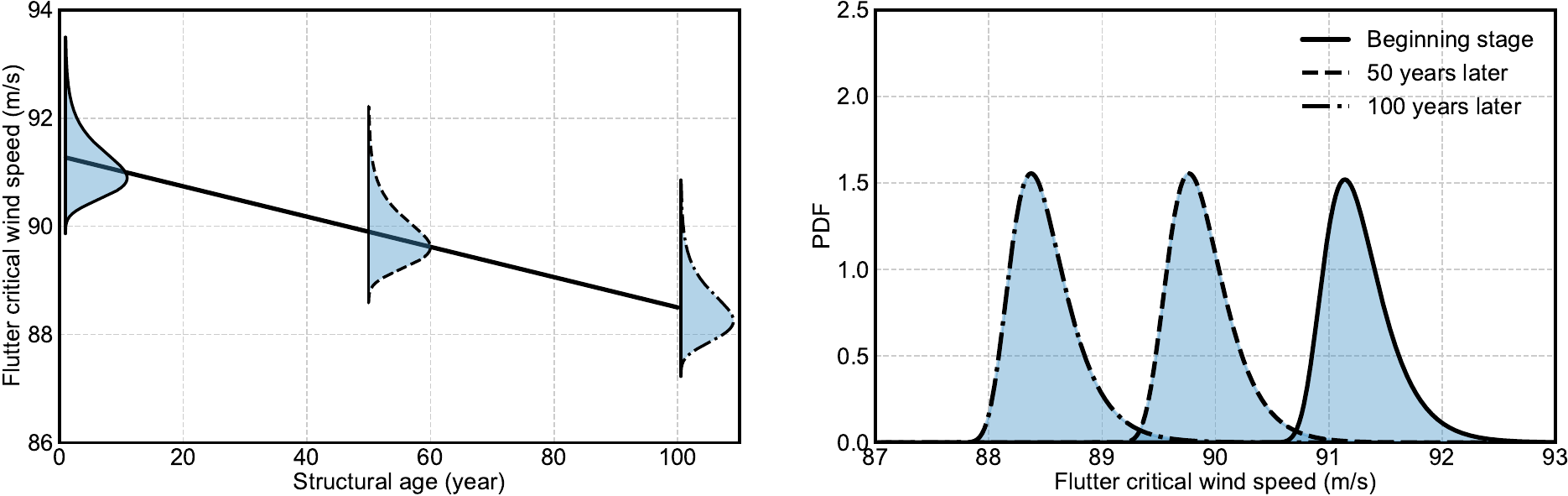}}
    \caption{Distributions of the flutter critical wind speed in futuristic 100-year structural age}
    \label{fig:flutter_year_distribution_moving}
\end{figure}

As shown in Fig.~\ref{fig:flutter_year_distribution_moving}, the mean value of the flutter critical wind speed in the long term tends to decrease in the structural life cycle, due to the deterioration effects of modal frequencies. Since only the mean values of modal frequencies' PDFs are changing and the standard variances of modal frequencies' PDFs are time-invariant, the standard variance of the flutter critical wind speed PDF will also be time-invariant.

\subsection{Life-cycle flutter probability evolution due to deteriorations of modal properties}
\label{sec:reliability_degradation_frequency_only}
\begin{figure}[!htb]
	\centering
	\includegraphics[]{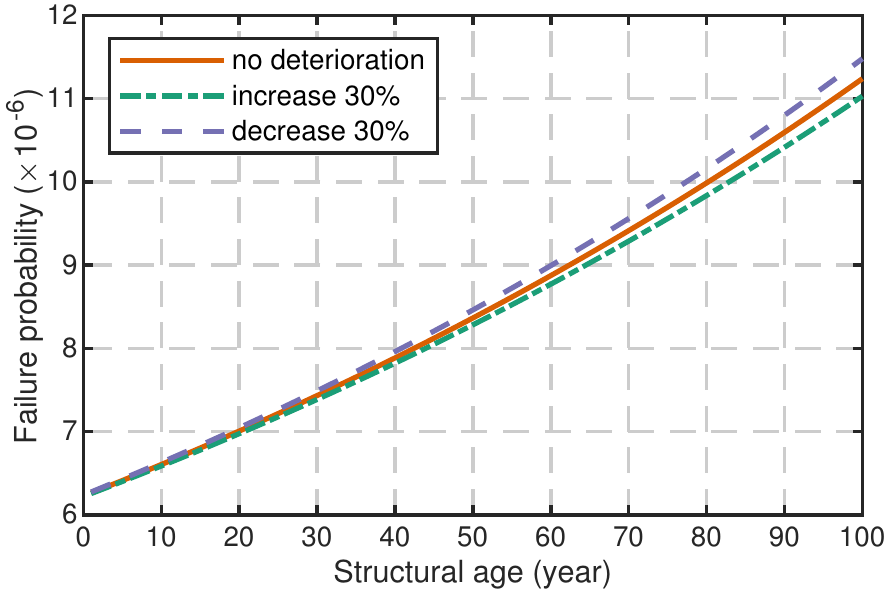}
	\caption{Comparison of various deterioration effects of damping ratios}
	\label{fig:failure_probability_extra}
\end{figure}

In this paper, the life-cycle flutter probability is calculated by Eq.~\eqref{eq:PIM}. Due to the time-invariant PDFs of damping ratios, the long-term flutter probability in the futuristic 100 years caused by deteriorations of modal frequencies is presented in Fig.~\ref{fig:failure_probability_extra} with the label ``no deterioration'', indicating that the failure probability might increase fairly fast with time, varying from $6.2\times10^{-6}$ to $11.2\times10^{-6}$.

Due to the limitation of the monitoring period (only available from 2010-2015), however, the long-term deterioration effect of damping ratios is vague. The changing trends of damping ratios will be clearer with a longer monitoring period. In order to clarify the potential deterioration effects of damping ratios, it is assumed that the mean values of damping ratios will increase or decrease by 30\% in 100 years later, rising or declining exponentially. The inter-seasonal fluctuations obey the same probability distributions mentioned in Sec~\ref{sec:damping_ratios_probability_distribution}. The long-term failure probability considering deterioration effects of modal frequencies and damping ratios simultaneously is illustrated in Fig.~\ref{fig:failure_probability_extra} with the labels ``increase 30\%'' and ``decrease 30\%''. 

As shown in Fig.~\ref{fig:failure_probability_extra}, deteriorations of damping ratios rarely affect the failure probability, which is plausible because even 30\% variation is still negligible compared with the inter-seasonal fluctuation. Therefore, it is suggested that more attention should be paid to the deteriorations of modal frequencies. 

\section{Conclusion} 
\label{sec:conclusion}
In this paper, an assessment methodology has been proposed to evaluate life-cycle flutter probability of long-span bridges based on field monitoring data. 
The 6-year dynamic properties including modal frequencies and damping rations are extracted by the field monitoring data and the fast Bayesian approach, where the deterioration effects and inter-seasonal fluctuations are discussed in detail.
Flutter critical wind speed is calculated by multi-mode flutter analysis based on the ideal structural model and the flutter derivatives identified by the wind tunnel test. A linear regression model is proposed to explicitly clarify how randomness is delivered from structural properties to the flutter critical wind speed, and to directly calculate the PDF of the flutter critical wind speed given the structural properties' PDFs. At last, the life-cycle flutter probability is calculated, considering the potential deteriorations of modal frequencies and damping ratios. Based on the application example of Xihoumen Bridge, the flutter probability might increase fairly fast with deteriorations of modal frequencies, but nearly remain the same with deteriorations of damping ratios. This study shows that the structural deteriorations, which are usually neglected for the flutter analysis, should be considered for the flutter-resistance design.

\section*{Acknowledgments}
The authors gratefully acknowledge the support of National Natural Science Foundation of China (52008314, 51778495, 51978527, 52078383) and Shanghai Pujiang Program (No.~19PJ1409800). Any opinions, findings and conclusions or recommendations are those of the authors and do not necessarily reflect the reviews of the above agencies.

\bibliography{wileyNJD-AMA.bib}%




\end{document}